\documentclass[twocolumn,prd,showpacs,preprintnumbers,aps]{revtex4}

\usepackage{graphicx}   

\newcommand{\vect}[1]{\mathbf{#1}}  

\newcommand{\mq}{|\vect{q}|}

\newcommand{\bbx}{\beta_{BX}}
\newcommand{\mmx}{\tilde{M}_X}
\newcommand{\mmb}{\tilde{M}_B}

\begin{document}

\preprint{OUTP-05-15P}
\preprint{hep-ph/0511169}

\title{Nonleptonic Weak Decays of B to $\mathbf{D_s}$ and D mesons}
\author{C. E. Thomas}
\email[E-mail: ]{c.thomas1@physics.ox.ac.uk}
\affiliation{Rudolf Peierls Centre for Theoretical Physics, University of Oxford,\\ 1 Keble Road, Oxford, OX1 3NP}
\date{23 March 2006}
\pacs{13.25.Hw, 12.15.Ji, 12.39.Jh}

\begin{abstract}
Branching ratios and polarization amplitudes for B decaying to all allowed pseudoscalar, vector, axial-vector, scalar and tensor combinations of $D_s$ and $D$ mesons are calculated in the Isgur Scora Grinstein Wise (ISGW) quark model after assuming factorization.  We find good agreement with other models in the literature and the limited experimental data and make predictions for as yet unseen decay modes.  Lattice QCD results in this area are very limited.  We make phenomenological observations on decays in to $D_s(2317)$ and $D_s(2460)$ and propose tests for determining the status and mixings of the axial mesons.  We use the same approach to calculate branching ratios and polarization fraction for decays in to two $D$ type mesons.
\end{abstract}

\maketitle

\section{Introduction}

Non-leptonic weak decays of B-mesons are important because they probe both electroweak physics and hadronic structure, and may provide a window on physics 
beyond the standard model.  As well as its intrinsic interest, the hadronic part must be understood in order to extract electroweak and new physics from these 
decays.  This involves non-perturbative QCD and so can not be calculated from first principles.  There have been many attempts to model the hadronic part and 
their success varies depending on which decay mode is studied.  There are only very limited lattice QCD results in this area.  Semi-leptonic decays of $B$ and 
$D$ mesons have been successfully studied using the Isgur Scora Grinstein Wise (ISGW) model \cite{ISGW}.  We extend this model to the study of exclusive 
non-leptonic decays after assuming factorization of these decays.  The model has corrections that vanish when the final mesons have zero recoil and so should be most reliable close to this kinematic region.  

Heavy Quark Effective Theory (HQET) as used, for example, in Ref.\ \cite{Rosner:HQET}, provides a set of symmetry relations but not expressions for the 
rates.  Some explicit model must still be used.  The ISGW model satisfies the requirements of HQET in the zero recoil limit \cite{ISGW2} and provides an explicit model in which calculations can be made.  

There are also pole models such as that of Bauer, Stech \& Wirbel (BSW) \cite{BSW:1}\cite{BSW:2} which we compare to our results.  There are different 
approaches for heavy to light decays such as the Soft Collinear Effective Theory (SCET) framework \cite{Bauer:2000yr} and Light-Cone Sum Rules 
\cite{Chernyak:LCSR}\cite{Belyaev:LCSR}.  These are useful where the final state mesons have high energy unlike the regime close to zero recoil that we are 
probing here.

Some predictions have been made for decays to combinations of pseudoscalar $D_s$ and $D$ and vector $D_s^*$ and $D^*$ such as by Luo \& Rosner 
\cite{Rosner:HQET} and Chen et. al. \cite{Chen:2005rp}.  Datta and O'Donnell \cite{Datta:2003re} have made limited studies of the decays to axial and scalar 
$D_s$.  Cheng et. al. in Ref. \cite{Cheng:2003sm} have studied s and p-wave form factors in the light-front approach.  They have compared this to improved 
ISGW model \cite{ISGW2} results given in Ref. \cite{Cheng:2003id}.  We specialize to the $D_s$ $D$ modes where the ISGW predictions should be reliable and 
calculate explicit results for all possible combinations of p and s-wave mesons consistently in one model giving polarization ratios where applicable.

With B factories tightening current observations of B decays and finding new decay modes it is timely that these modes should be studied in detail.  

Our aims are to:
\begin{itemize}
\item Provide robust predictions of branching ratios and polarization fractions for all the possible combinations.  These can be compared with current and future experimental data.
\item Comment on what can be learnt about the nature of $D_s(2317)$ and $D_s(2460)$ mesons and axial vector mixing.
\end{itemize}

We start in Section \ref{sec:General} with a discussion of factorization and general remarks.  In Section \ref{sec:ISGW} we briefly describe the ISGW model and extend it to non-leptonic decays after assuming factorization.  In Section \ref{sec:Calculation} we present our results for the branching ratios and polarization fractions of the allowed 
decay modes.  We also compare our results with those from other models and experimental data in this section.  Then we apply the same method to decays in to two $D$ type mesons.  In Section \ref{sec:PhenomDs0Ds1} we discuss what we can learn about the scalar and axial $D_{s0}$ and $D_{s1}$.  We finish with some general remarks and conclusions in Section \ref{sec:Conclusions} and suggest some experimental observations that could be made.

\section{General Definitions and Factorization}
\label{sec:General}

\begin{figure}[tb]
\begin{center}
\includegraphics[width=8cm]{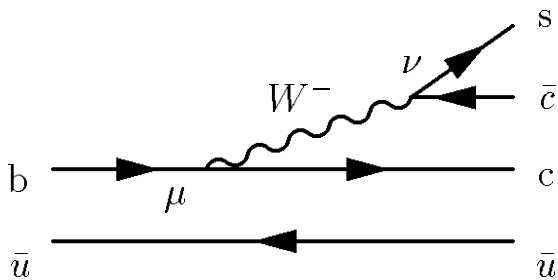}
\caption{$B^- (b\bar{u}) \rightarrow D_s^- (\bar{c} s)  D^0 (c \bar{u})$ - Type I Tree Diagram}
\label{fig:TypeITree}
\end{center}
\end{figure}

The color favored tree diagram (``Type I tree diagram'') for $B \rightarrow D_s D$ is shown in Fig.\ \ref{fig:TypeITree}.  Here $D_s$ and $D$ can be any s or p-wave $s\bar{c}$ and $c\bar{u}$ states.  We ignore any contribution to this process from penguin and weak-annihilation topologies.

The rate for a general Type I Tree decay $B(b\bar{q_i}) \rightarrow Y(q_1 \bar{q_2}) X(q \bar{q_i})$ (e.g. $Y = D_s$ and $X = D$) can be written as:
\begin{equation}
\Gamma = \frac{G_F^2}{16 \pi} a_1^2 |V_{q b} V_{q_1 q_2}|^2 \frac{\mq}{M_B^2} |A|^2 .
\end{equation} 
where $G_F$ is the Fermi Constant, $a_1$ is the effective Wilson coefficient, $V_{q b}$ and $V_{q_1 q_2}$ are Cabibbo-Kobayashi-Maskawa (CKM) quark-mixing matrix elements, $\vect{q}$ is the recoil 3-momentum in the rest frame of $B$ and $M_B$ is the B-meson mass.  $|A|^2$ is the sum of the squares of the polarization amplitudes $A_i$:
\begin{equation}
|A|^2 \equiv \sum_{i}^{} |A_i|^2 .
\end{equation}
We use the notation $i$ = $+-$, $-+$ or $ll$ where the first label denotes the helicity of the $X$ meson.

We follow Rosner \& Luo \cite{Rosner:HQET} by assuming the factorization hypothesis (naive factorization) and writing the amplitude as a product of two matrix elements:
\begin{equation}
A_i = <Y|J_{\mu}|0><X|J^{\mu}|B> ,
\end{equation}
where $J_{\mu} \equiv V_{\mu} - A_{\mu} \equiv \bar{q}_f\gamma_{\mu}(1-\gamma_5)q_i$ is the vector-axial current.  In QCD Factorization \cite{BBNS:1999} \cite{BBNS:2000} \cite{BN:2003} it has been shown that B decays to two heavy mesons do not obey the BBNS factorization formula and so naive factorization is not expected to hold.  However, as noted in Ref. \cite{BBNS:2000}, the charm quark mass is at an intermediate scale between the light and heavy (bottom quark mass) scales and so factorization is still an open question in the modes we consider.  Empirically factorization works where it has been tested in these modes \cite{Rosner:HQET} \cite{Bortoletto:1990np} \cite{Chen:2005rp}.  Comparison of our results with experiment can further test this assumption.  The effective Wilson coefficient $a_1$ contains some QCD corrections.  As remarked in Ref. \cite{Rosner:HQET} this varies from process to process but only by less than about 1\%.  We therefore use that paper's value, $a_1 =  1.05$, for all processes. 
	
The matrix element $<X|J^{\mu}|B>$ can be parameterized in terms of general form factors.  We use the same definitions as ISGW and these are given in Appendix 
\ref{app:GeneralV-A}.  These form factors can then be calculated in a particular model such as ISGW discussed in Section \ref{sec:ISGW}.  The polarization 
vectors and tensors used for $J=1$ and $J=2$ mesons are given in Appendix \ref{app:PolVectors}. 

The matrix element $<Y|J_{\mu}|0>$ can be parameterized in terms of the decay constant of meson Y, $f_Y$.  For vector ($^3S_1$) and axial vector ($^3P_1$, 
$^1P_1$) mesons \footnote{There are many definitions in the literature.  Note that here we use the same conventions for vector and axial vector mesons.} with 
polarization vector $\epsilon_{\mu}^{Y}$ and mass $M_Y$ this is defined by
\begin{equation}
<Y|J_{\mu}|0> = \epsilon_{\mu}^{*Y} f_Y M_Y ,
\end{equation}
and for scalar ($^3P_0$) and pseudoscalar ($^1S_0$) mesons \footnote{Again there are many different conventions here.  We use the same convention for scalar 
and pseudoscalar mesons.} with 4-momentum $P_{\mu}^{Y}$,
\begin{equation}
<Y|J_{\mu}|0> = i P_{\mu}^{Y} f_Y .
\end{equation}

\section{The ISGW model}
\label{sec:ISGW}

The ISGW model \cite{ISGW} was developed to calculate semileptonic $B$ and $D$ decays using a constituent quark model and the mock meson approach 
\cite{Hayne:1981zy}.  We extend the model to non-leptonic decays using the factorization hypothesis and test the sensitivity to the assumptions made.  We find that robust predictions can be 
made for the $D_s$ $D$ decay modes because they are in the valid kinematic region.

The matrix element $<X|J^{\mu}|B>$ is calculated using a non-relativistic decomposition of the quark current $\bar{q} \gamma_{\mu}(1-\gamma_5) b$.  This will 
contain corrections of $O(\mq^4 / m_{q_i}^4)$ and  $O(\mq^4 / M_X^4)$ where $m_{q_i}$ are the quark masses appearing in the current decomposition.  The model 
is therefore reliable only close to zero recoil and with heavy quarks.  
$M_B$ is the mass of the B meson, $M_X$ is the mass of the D (or excited D) meson, $M_Y$ is the mass of the $D_s$ (or excited $D_s$) meson.  $\tilde{M}_B$, 
$\tilde{M}_X$ and $\tilde{M}_Y$ are the respective mock meson masses.  $P_B$, $P_X$, $P_Y$ are the 4-momenta,  $y \equiv \frac{t}{M_B^2} \equiv \frac{(P_B - 
P_X)^2}{M_B^2} = \frac{M_Y^2}{M_B^2}$, and $t_m \equiv \max{t} = (M_B - M_X)^2$.  The quarks are the spectator antiquark $\bar{q}_i$, the decaying quark $b$, 
the quark in the final state meson with the spectator quark $q$, and the quark and antiquark pair in the other final state meson $q_1$ and $\bar{q}_2$.  
$m_{qi}$, $m_b$, $m_q$, $m_{q1}$ and $m_{q2}$ are the constituent quark masses.  In the decay $B^- \rightarrow D_s^- D^0$: $\bar{q}_i = \bar{u}$, $q = c$, 
$q_1 = s$ and $\bar{q}_2 = \bar{c}$.

In the original ISGW paper the semileptonic differential rate is given in terms of the form factors.  We have adapted the model to calculate non-leptonic 
decays in the factorization hypothesis given above.  The resulting relationships between polarization amplitudes, form factors and decay constants are given 
in Appendix \ref{app:PolAmplitudes}.  For connection with the original ISGW paper \cite{ISGW}, their hadronic tensor (Equ.\ 7) is given by
\begin{equation}
h_{\mu\nu} = \sum_{i}^{} <B|J_{\nu}^{\dagger}|X> <X|J_{\mu}|B> .
\end{equation}

We initially followed the original ISGW paper using (radial ground state) harmonic oscillator (HO) wavefunction of the momentum space form:
\begin{eqnarray}
\psi_{l=0,m_l=0}(\vect{k}) = \left(\pi\beta^2\right)^{-3/4} \exp{\left(-\frac{|\vect{k}|^2}{2\beta^2}\right)} ,\\
\psi_{l=1,m_l=0(1)}(\vect{k}) = \sqrt{2} \left(\pi\beta^2\right)^{-3/4} \frac{k_{z(+)}}{\beta} \exp{\left(-\frac{|\vect{k}|^2}{2\beta^2}\right)} .
\end{eqnarray}
where $\beta$ is the wavefunction parameter and $k_+ = - \frac{1}{\sqrt{2}}(k_x + i k_y)$ .  The wavefunctions are normalized such that $\int d^3k \psi^*(k) 
\psi(k) = 1$.

The resulting ISGW form factors using harmonic oscillator (HO) wavefunctions are given in Appendix \ref{app:ISGWFormFactors}.  These are mostly the same as 
the ISGW paper with some additional ones not given in that paper.  However, there are a couple of differences.  The sign of the scalar form factor $u_+$ in Equ.\ \ref{Equ:ISGWFormFactor:up} is opposite to that in Equ.\ B37 of Ref.\ \cite{ISGW}.  As Ref.\ \cite{ISGW} does not give an expression for $u_-$ it is not possible to tell if this is an overall or relative sign.  It seems likely that it is an overall sign due to the conventions used, in which case it has no affect on the results.  If it is a relative sign it can only modify the results when meson $Y$ is a pseudoscalar or scalar.  If, for example, $Y$ is the pseudoscalar $D_s$, adding a relative sign changes the branching ratio significantly: from $4.2 \times 10^{-4}$ to $2.2 \times 10^{-3}$.

The other difference is the additional mass ratio $\mmb/\mmx \approx 2.2$ in the axial vector ($1^{+-}$) form factor $v$ in Equ.\ B43 of Ref.\ \cite{ISGW} compared with Equ.\ \ref{Equ:ISGWFormFactor:v} here.  This is a parity violating form factor which appears in decays where meson $Y$ is a vector or an axial vector and only in the transverse polarization amplitudes.  This has a negligible affect on the branching ratios and longitudinal fractions.  For example, when $Y$ is the vector $D_s^*$, the branching ratios (arbitrary normalization) are $\propto 5.28 \times 10^{13}$ and $5.29 \times 10^{13}$, and the longitudinal fractions $0.930$ and $0.928$ respectively without and with the mass ratio.  However, there is a significant affect on the parity odd fraction $\propto (A_{+-} - A_{-+})^2$.  This is $0.66 \times 10^{-3}$ without the mass ratio but $3.1 \times 10^{-3}$ with it.

The original ISGW paper was restricted to harmonic oscillator wavefunctions; our ISGW form factors for general wavefunctions are given in Appendix \ref{app:ISGWGenFormFactors}.

In the original ISGW paper $\mq^2$ in form factors is approximated as $(t_m - t)M_X/M_B$ and a relativistic correction factor $\kappa$ is introduced through 
$\mq^2 \rightarrow \mq^2/{\kappa^2}$.  This was intended to bring the pion form factor into better agreement with experiment at the low $q^2$ region 
\cite{ISGW}.  We take $\kappa = 0.75$ \cite{JoThesis}.  However, we look at how much $\kappa$ influences the results by setting $\kappa = 1$ in one model.

There is ambiguity in the mock meson approach \cite{Capstick:1989ra} as to whether, at the end of the calculation, to identify the mock meson masses as the 
physical meson masses or the sum of the constituent quark masses.  

We used a number of different models corresponding to different choices to assess the robustness of the approximations used.  In model 1 the $(t_m - t)$ form 
is used in place of $\mq^2$.  In models 2 and 3 the exact $\mq^2$ is used in the exponential of $F_3$.  In models 1-2 the mock meson masses are equal to the 
physical meson masses.  In model 3 the physical meson masses are used except in the exponential of $F_3$ where the sum of the constituent quark masses is 
used.  Models 4-5 take the mock meson masses to be the sum of the constituent quark masses.  Model 4 uses $(t_m - t)$ and model 5 uses $\mq^2$.  Model 6 is 
the same as model 3 except that we set $\kappa = 1$.

The $D_s$ $D$ decay mode is expected to be reasonably well modeled by ISGW.  The highest recoil momentum (that for final state pseudoscalars) is 1812 MeV.  
$\mq / m_b \approx 0.35$, $\mq / m_c \approx 1.02$, $\mq / M_D \approx 0.97$.  Although $\mq / m_c$ and $\mq / M_D$ do not appear small, they appear as, for 
example, $\mq^4 / (8 M_D^4)$ in a Taylor expansion with numerical value $\approx 0.14$.  The approximation is therefore reasonable.  In 
the following section we present the results of our calculation.

\section{Results of Calculations}
\label{sec:Calculation}

\subsection{ISGW Model Results}
\label{sec:ISGWModelResults}

\begin{table}[tb]
\begin{center}
\begin{tabular}{|c|c|}
\hline
\textbf{Quark} & \textbf{Constituent Quark Mass / MeV} \\
\hline
$b$ & 5170 \\
\hline
$q = q_1 = c$ & 1770 \\
\hline
$q_i = u$ & 330 \\
\hline
$q_2 = s$ & 550 \\
\hline
\end{tabular}
\end{center}
\caption{Constituent quark masses \protect\cite{JoThesis}.}
\label{table:quarkmasses}
\end{table}

\begin{table}[tb]
\begin{center}
\begin{tabular}{|c|c|}
\hline
\textbf{Meson} & \textbf{ $\beta$ / MeV} \\
\hline
$B$ & 410 \\
\hline
$D$ & 390 \\
\hline
$D^*$ & 390 \\
\hline
$D_0$ & 330 \\
\hline
$D_1$ & 330 \\
\hline
$D_2$ & 330 \\
\hline
\end{tabular}
\end{center}
\caption{HO wavefunction parameters \protect\cite{CloseDudek:HybridEMWeak}.}
\label{table:wavefunctionparams}
\end{table}

\begin{table}[tb]
\begin{center}
\begin{tabular}{|c|c|}
\hline
\textbf{Meson} & \textbf{Decay Constant $f$ / MeV} \\
\hline
$D_s$ & 240 MeV \cite{NS} \\
\hline
$D_s^*$ & 275 MeV \cite{NS} \\
\hline
$D_{s0}$ & 110 MeV \cite{Veseli:DecayConstants} \\
\hline
$D_{s1}1$ & 240 MeV \cite{Veseli:DecayConstants} \\
\hline 
$D_{s1}2$ & 63 MeV \cite{Veseli:DecayConstants} \\
\hline
\end{tabular}
\end{center}
\caption{$D_s$ meson decay constants.}
\label{table:DecayConstantsDsD}
\end{table}

Unless otherwise stated all numerical values of constants, masses etc. are taken from the PDG Review 2004 \cite{PDG04}.  The quark masses and HO wavefunction 
parameters used are shown in Tables \ref{table:quarkmasses} and \ref{table:wavefunctionparams} respectively.  We take $a_1$ = 1.05 \cite{Rosner:HQET}, 
$|V_{bc}|$ = 0.04 and $|V_{cs}|$ = 0.97.  The magnitudes of decay constants used are given in Table \ref{table:DecayConstantsDsD}.  The phases of the decay 
constants are chosen to match those given by quark model calculations in Appendix \ref{app:DecayConstants}.

To be explicit, we consider the mode in which a charged $B$ decays in to a charged $D_s$ and neutral $D^0$.  Initially we take the axial vector $D_{s1}(2460) \equiv D_{s1}1$ to be $^3P_1$ with mass 2460 MeV.  The $D_{s1}(2536) \equiv D_{s1}2$ is $^1P_1$ and has mass 2536 MeV.  We discuss axial vector mixing in Section \ref{sec:PhenomDs0Ds1}.  The scalar $D_{s0}$ is assumed to be the $D_{s}(2317)$ with mass 2317 MeV.  We take both axial $D_{1}$ to have mass 2420 MeV: $D_{1}1$ is $^3P_1$ and $D_{1}2$ is $^1P_1$.  The scalar $D_{0}$ has mass 2350 MeV and the tensor $D_{2}$ has mass 2459 MeV.

As discussed in Section \ref{sec:ISGW}, to get a handle on the theoretical uncertainty within the ISGW model, we calculated the branching ratios and 
polarization ratios using six different model choices.

A plot of the calculated decay rates and the experimental results is shown in Fig.\ \ref{fig:DsDPlot}.  The model error bars and average show the range and 
average of the six model choices.  The experimental results are from the PDG Review 2004 \cite{PDG04} apart from: $B^0 \rightarrow D_s^+ D^-$ which is from 
Belle \cite{Belle:B0DsD2005} and $B^0 \rightarrow D_s^{*+} D^{*-}$ which is from BABAR \cite{Aubert:2005xu}.  Tables of numerical results are given in Tables \ref{table:DsDISGWBRs} and \ref{table:DsDISGWPolFractions} of Appendix \ref{app:NumericalResults}.  Products of branching ratios have been measured for the $D_s(2317)$ and $D_s(2460)$ modes from BABAR \cite{Aubert:2004pw} and Belle \cite{Krokovny:2003zq}.  This data is averaged in the PDG Review 2005 Partial Update \cite{PDG05} and discussed in Section 
\ref{sec:PhenomDs0Ds1}.

Where data exist the model seems in remarkable agreement.  The possible exceptions are the newer Belle and BABAR results for the neutral $B$ decaying to 
$D_s^- D^+$ and $D_s^{*-} D^{+*}$.  Both these neutral $B$ modes are lower than the model predictions and, although this is not significant, they are lower 
than the corresponding charged $B$ mode.  If more precise data shows a discrepancy between charged and neutral decays it would be due to some difference 
between $u$ and $d$ spectator quarks.  Spectator interaction effects which could cause this include weak annihilation, which can only contribute to the charged $B^+$ decay, and electromagnetic penguins.

The variation due to the model choice is relatively small; it's of the same order or less than the experimental uncertainty for those modes with experimental 
data.  This suggests the predictions made are robust.  The model uncertainty shown does not include uncertainty in the decay constants: this would change the 
overall branching ratio for each $Y$ (e.g. $D_s$) meson but not the pattern for a given $Y$.

The $D_{s1}1$ $D^*$, $D_{s1}1$ $D$ and $D_{s1}2$ $D^*$ modes are relatively large but are complicated by axial vector mixing.  We comment on implications for 
$D_{s0}$ and $D_{s1}$ mesons in Section \ref{sec:PhenomDs0Ds1}.

Polarization ratios for the relevant decay modes are shown in Fig.\ \ref{fig:DsDPolarisations}.  There is only one piece of experimental data in the PDG Review 2004: that from BABAR and CLEO on the decay in to $D_s^*$ $D^*$.  Here there is good model agreement.  The model uncertainty is $\sim 10\%$ being 
slightly larger for the $D_11$ modes.  Again the model predictions are robust.

\begin{widetext}

\begin{figure}[tb]
\begin{center}
\includegraphics[width=11cm,angle=270]{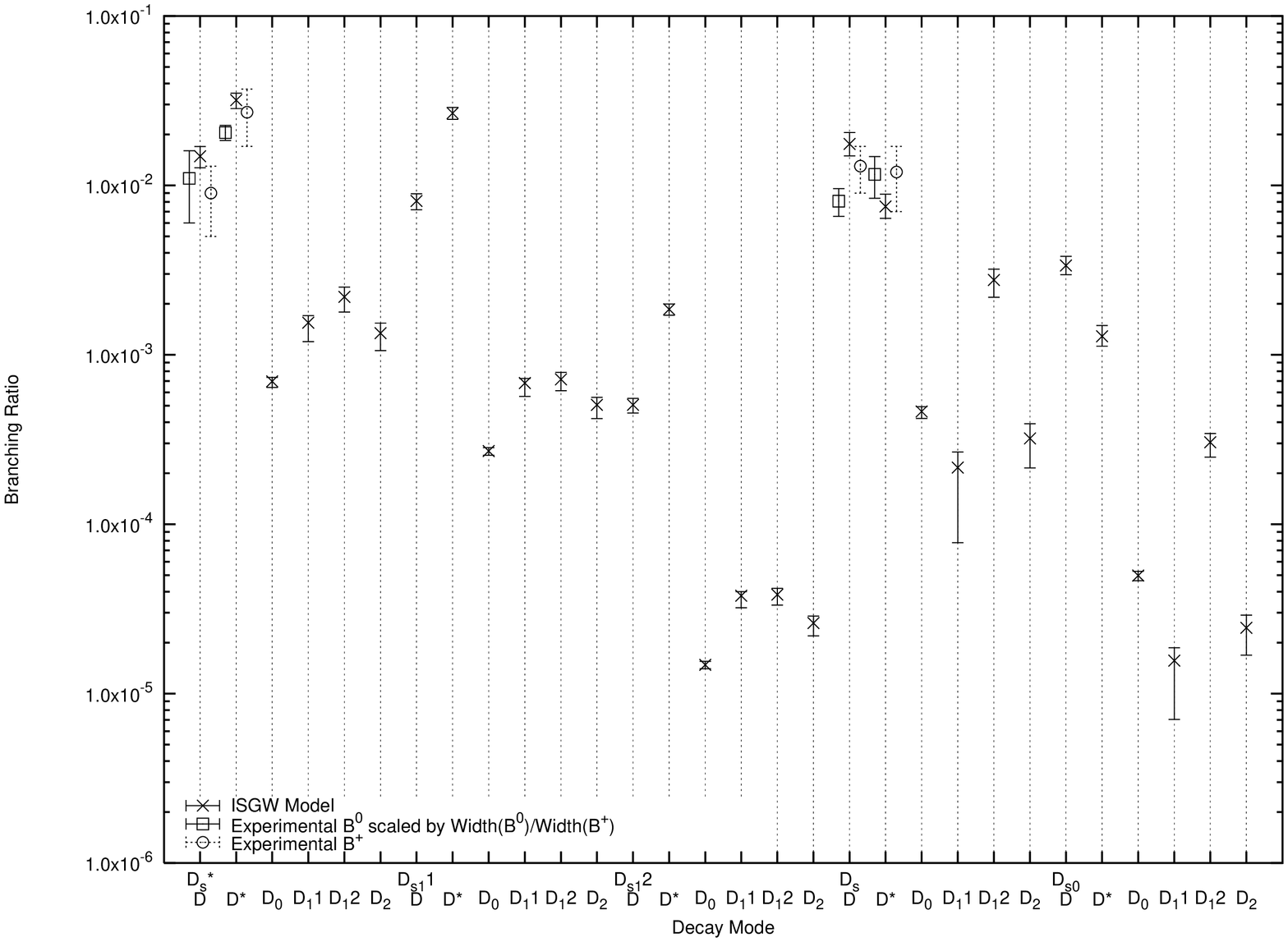}
\caption{$B \rightarrow D_s D$ Branching Ratios: ISGW model and experimental values.  See the text for references to experimental results.}
\label{fig:DsDPlot}
\end{center}
\end{figure}

\begin{figure}[tb]
\begin{center}
\includegraphics[width=11cm,angle=270]{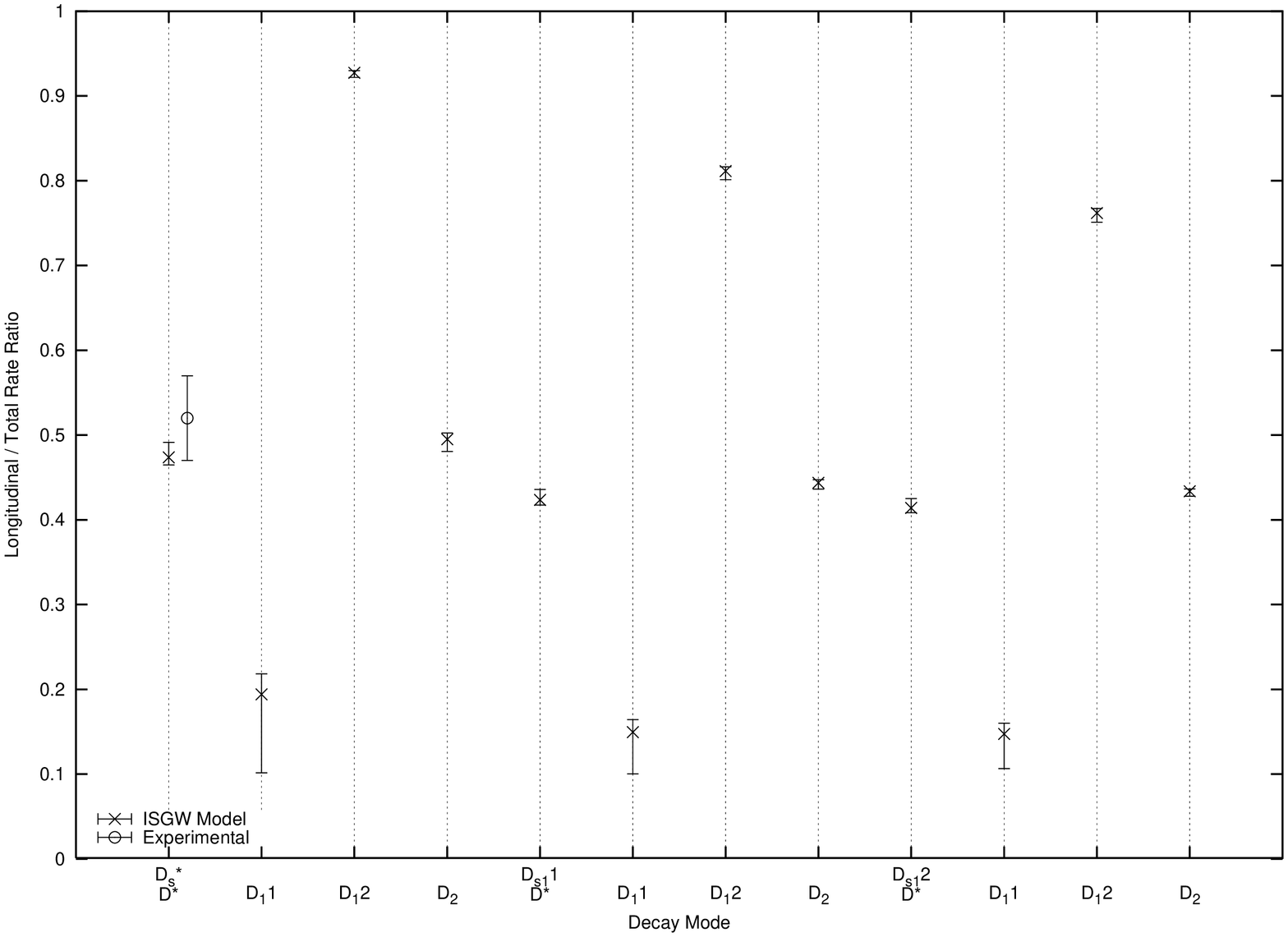}
\caption{Polarization Ratios: ISGW model and experimental values.  See the text for references to experimental results.}
\label{fig:DsDPolarisations}
\end{center}
\end{figure}

\end{widetext}

We then looked at the ISGW model with more realistic wavefunctions using the results of Appendix \ref{app:ISGWGenFormFactors} and a numerical solution of the 
radial Schr\"{o}dinger equation.  This was done by discretizing the position coordinate in to $300$ intervals ranging from $0$ to $\approx 2 \times 10^{-2} 
\text{MeV}^{-1}$.  We first checked our numerical method by solving a HO potential $V = 1/2 \mu \omega^2 r^2$ where $\omega = \beta^2 / \mu$ and $\mu$ is the 
reduced mass of the quarks.  The results were the same as the HO analytic ones to the numerical accuracy.  

We then used a more realistic Linear + Coulomb + Hyperfine (LCHF) potential with the delta function smoothed out to a Gaussian 
\cite{Swanson:1992ec}\cite{ESwanson:numericalsolution}:
\begin{equation}
V = - \frac{4}{3}\frac{\alpha_s}{r} + b r + \frac{32 \alpha_H \sigma^3}{9 m_q m_{\bar{q}} \sqrt{\pi}}\left(\vect{s_q}\cdot \vect{s_{\bar{q}}}\right) 
\exp{(-\sigma^2 r^2)} .
\end{equation}
Taking $\alpha_s = 0.594$, $b = 1.62 \times 10^5 \text{MeV}^2$, $\sigma = 897 \text{MeV}$ and $\alpha_H = \alpha_s$.  The resulting form factors are 
negligibly different from the HO ones compared with the other model uncertainties.  Varying the parameters in the LCHF potential by $10\%$ leads to negligible 
changes in the results.

So, to summarize, the potential used does not appear to modify our results significantly.

\begin{widetext}

\begin{figure}[tb]
\begin{center}
\includegraphics[width=9cm,angle=270]{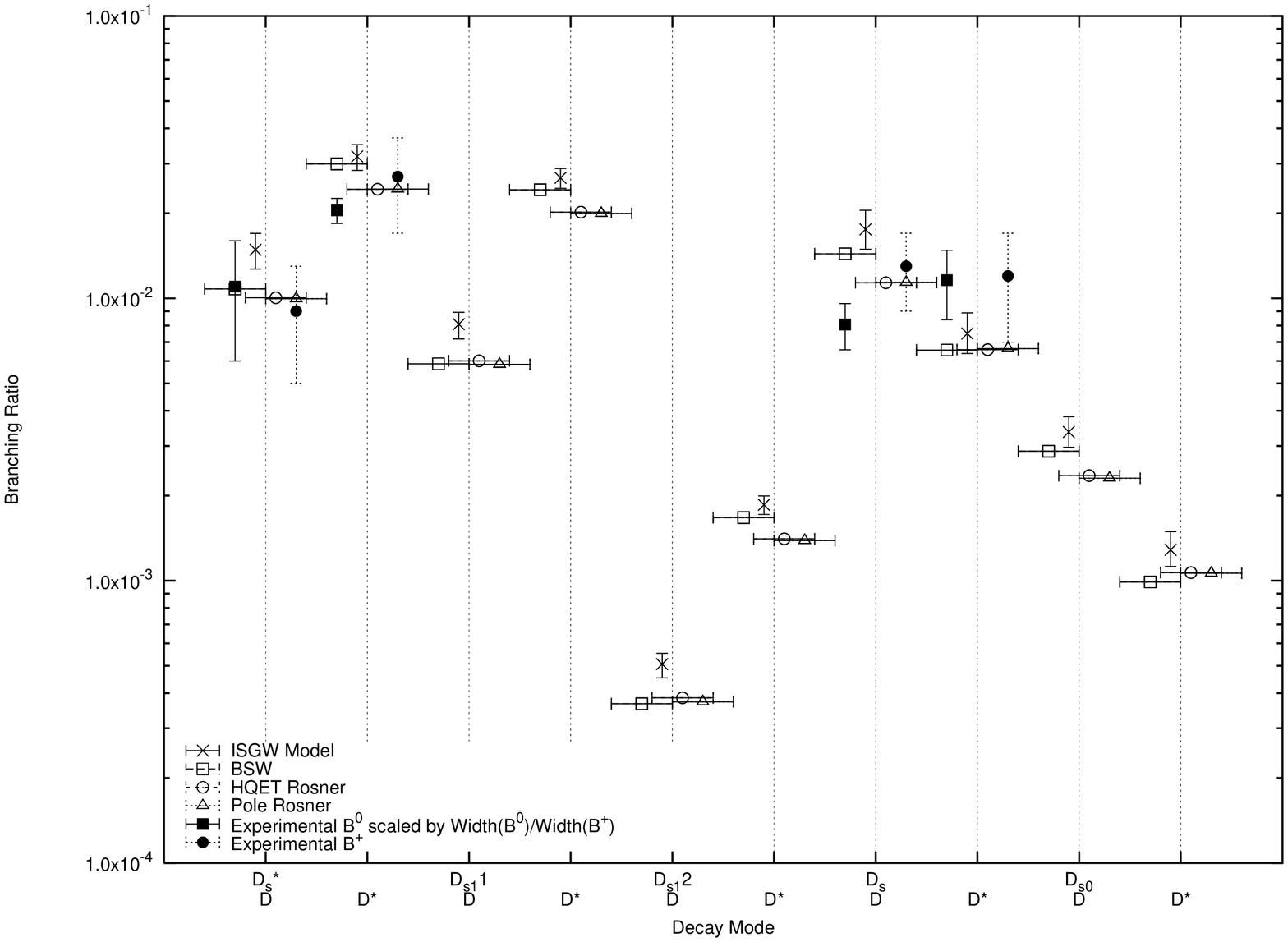}
\caption{$B \rightarrow D_s D$ Branching Ratios: Different models.  Experimental data as in Fig.\ \protect\ref{fig:DsDPlot}.}
\label{fig:DsDPlot_others}
\end{center}
\end{figure}

\begin{figure}[tb]
\begin{center}
\includegraphics[width=8cm,angle=270]{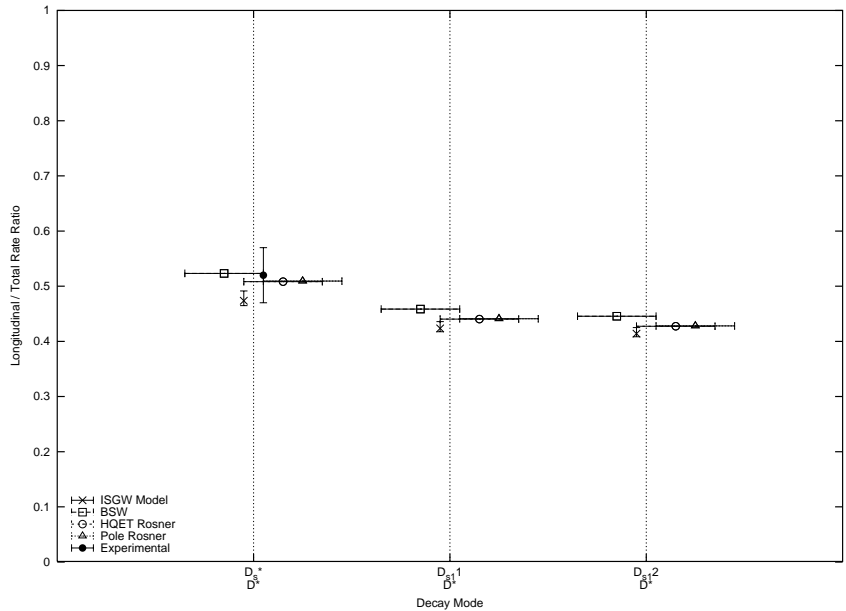}
\caption{Polarization Ratios: Different models.  Experimental data as in Fig.\ \protect\ref{fig:DsDPolarisations}.}
\label{fig:DsDPolarisations_others}
\end{center}
\end{figure}

\end{widetext}

\subsection{Comparison with Other Results}
\label{sec:Comparison}

We compared these results with calculations based on the BSW model \cite{BSW:1}\cite{BSW:2}, Luo and Rosner's HQET based model in ref. \cite{Rosner:HQET} and 
the pole model in that reference.  We used our values of the masses, decay constants, CKM matrix elements and $a_1$. These models only allow predictions when 
meson X is a vector or pseudoscalar meson.  The branching fraction results are shown in Fig.\ \ref{fig:DsDPlot_others} and the polarization ratios in Fig. 
\ref{fig:DsDPolarisations_others}.  It can be seen that there is reasonably good agreement between the four models.  This adds to our confidence in the 
robustness of our results.  The ISGW model appears to give a slightly worse fit in some cases.  However we should note that we haven't fitted anything to these decays.  The only parameter, $a_1$, was obtained for general $B$ decays. 

Cheng et. al. \cite{Cheng:2003sm} have used a light-front approach to calculate some of these decay modes.  They do not, however, give explicit results for all the modes that can be compared to experimental data.  The light front approach has the advantage of being relativistic, however, in the region close to zero recoil being studied we have argued that the ISGW model should be valid.  The light front approach uses a parameterization to get from maximum recoil to the physical regime.

Cheng \cite{Cheng:2003id} has also studied s and p-wave form factors in an improved ISGW model \cite{ISGW2}.  However different models are used for different transitions.  For consistency we have used the same model for all the decay modes.  We give explicit results for all possible combinations including p-wave and and p-wave modes some of which are comparable in magnitude to the p-wave + s-wave modes.  We present polarization ratios where appropriate.  None of the parameters have been fit to these decays; the decay constants are determined from other experimental results or models.  We have also attempted to assess the impact of some of the approximations in our results.  Their qualitative pattern is the same as ours but there are quantitative differences, comparison being complicated by mixing effects.

Lattice calculations currently only determine the form factors at zero recoil for semileptonic decays of $B$ to $D$ or $D^*$.  Therefore branching ratios can 
not be compared with our calculations.  For $B \rightarrow D$ the lattice results are given as $F(w=1)$ where $w = P_B \cdot P_X / (M_X M_B)$.  This is 
related to the ISGW form factor $f_+$ by
\begin{equation}
F = \frac{2 \sqrt{M_X M_B}}{M_X + M_B} f_+ .
\end{equation}
The ISGW model gives $0.961$ to $1.02$ depending on what masses are used for the mock meson masses.  Lattice calculations give $1.074\pm0.018\pm0.016$ 
\cite{Okamoto:2005zg}.  For $B \rightarrow D^*$ the lattice results are given as $F_A(w=1)$.  This is related to the ISGW form factor $f$ by 
\begin{equation}
f = \sqrt{M_X M_B} F_A(w) (1 + w) .
\end{equation}
The ISGW model gives this as $0.998$ whereas a quenched lattice calculation gives $0.913^{+0.024}_{-0.017}\pm0.016^{+0.003+0.000+0.006}_{-0.014-0.016-0.014}$ 
\cite{Hashimoto:2001nb}.  In both cases the results of lattice and ISGW models differ.

\subsection{The $B \rightarrow D D$ Modes}
\label{sec:B_DDResults}

The ISGW model can also be used in decays to two $D$ (and excited) mesons.  The dominant contribution is again expected to be from the Type I tree diagram 
shown in Fig.\ \ref{fig:TypeITree}.  In the decay $B^- \rightarrow D^- D^0$: $\bar{q}_i = \bar{u}$, $q = c$, $q_1 = d$ and $\bar{q}_2 = \bar{c}$.  These decays are Cabibbo suppressed with the CKM matrix element $V_{cs}$ replaced by $V_{cd}$ leading to a suppression factor of $0.05$ in rate compared with the $D_s$ $D$ modes.  This means that other contributions may be relatively more important.  

An additional complication is in distinguishing the two $D$ mesons produced.  In neutral $B$ decays the $B^0$ and $\bar{B^0}$ decays must be distinguished to 
determine which $D$ meson is which.  In the $B^0$ decay meson $Y$ has charge $+$ whereas in the $\bar{B^0}$ decay it has charge $-$.  This is not a problem in 
charged $B$ decays where meson $Y$ is always charged and $X$ is neutral.

With these caveats the decays can be calculated as above.  We take $|V_{cd}|$ = 0.22.  The magnitudes of decay constants used are given in Table 
\ref{table:DecayConstantsDD}.  These are from a quark model calculation using HO wavefunctions, as described in Appendix \ref{app:DecayConstants}, normalized 
to the $f_D$ result from CLEO \cite{Artuso:2005ym}.  They produce the same pattern of results as in Ref. \cite{Veseli:DecayConstants}. 

\begin{table}[tbh]
\begin{center}
\begin{tabular}{|c|c|}
\hline
\textbf{Meson} & \textbf{Decay Constant $f$ / MeV} \\
\hline
$D$ & 223 \cite{Artuso:2005ym}  \\
\hline
$D^*$ & 215 \\
\hline
$D_{0}$ & 156 \\
\hline
$D_{1}1$ & 183 \\
\hline 
$D_{1}2$ &  89 \\
\hline
\end{tabular}
\end{center}
\caption{$D$ meson decay constants}
\label{table:DecayConstantsDD}
\end{table}

The branching ratios and polarization fractions are shown Figures \ref{fig:DDPlot} and \ref{fig:DDPolarisations} respectively.  Numerical values are given in 
Tables \ref{table:DDISGWBRs} and \ref{table:DDISGWPolFractions} of Appendix \ref{app:NumericalResults}.  In the PDG Review 2005 Update \cite{PDG05} there are 
only upper limits for the vector and pseudoscalar modes along with a combined branching ratio in to $D^+$ $D^*$ and $D^{*+}$ $D$ of $0.93 \pm 0.15 \times 
10^{-3}$.  The experimental data shown on the figures is from Belle \cite{Miyake:2005qb} \cite{Majumder:2005kp}.  There is good agreement except for decay to $D$ $D$.  Again here the $B^0$ branching ratio is lower than the $B^+$ branching ratio.  The model uncertainty shown does not include 
uncertainty in the decay constants which could be significant.

Because of the charge conjugation symmetry of the final state in decays to $D^*$ $\bar{D}^*$ another interesting observable is the CP odd fraction, $R_{\perp}$.  
This is related to the helicity amplitudes: Equs.\ \ref{eqn:polamps:vv:pm} and \ref{eqn:polamps:vv:mp} in Appendix \ref{app:PolAmplitudes} by
\begin{equation}
R_{\perp} = \frac{|A_{+-}-A_{-+}|^2}{2(|A_{+-}|^2 + |A_{-+}|^2 + |A_{ll}|^2)}
\end{equation}
The model $R_{\perp}$ are shown in Table \ref{table:DDISGWTransFractions} of Appendix \ref{app:NumericalResults}.  There is experimental data on $R_{\perp}$ in the decay to $D^* D^*$: Belle gives $R_{\perp} = 0.19 \pm 0.08 \pm 0.01$ \cite{Miyake:2005qb} and BABAR $R_{\perp} = 0.125 \pm 0.044 \pm 0.007$ 
\cite{Aubert:2005rn}.  This is in general agreement with $R_{\perp}=0.086$ from the ISGW model.

\begin{widetext}

\begin{figure}[tb]
\begin{center}
\includegraphics[width=11cm,angle=270]{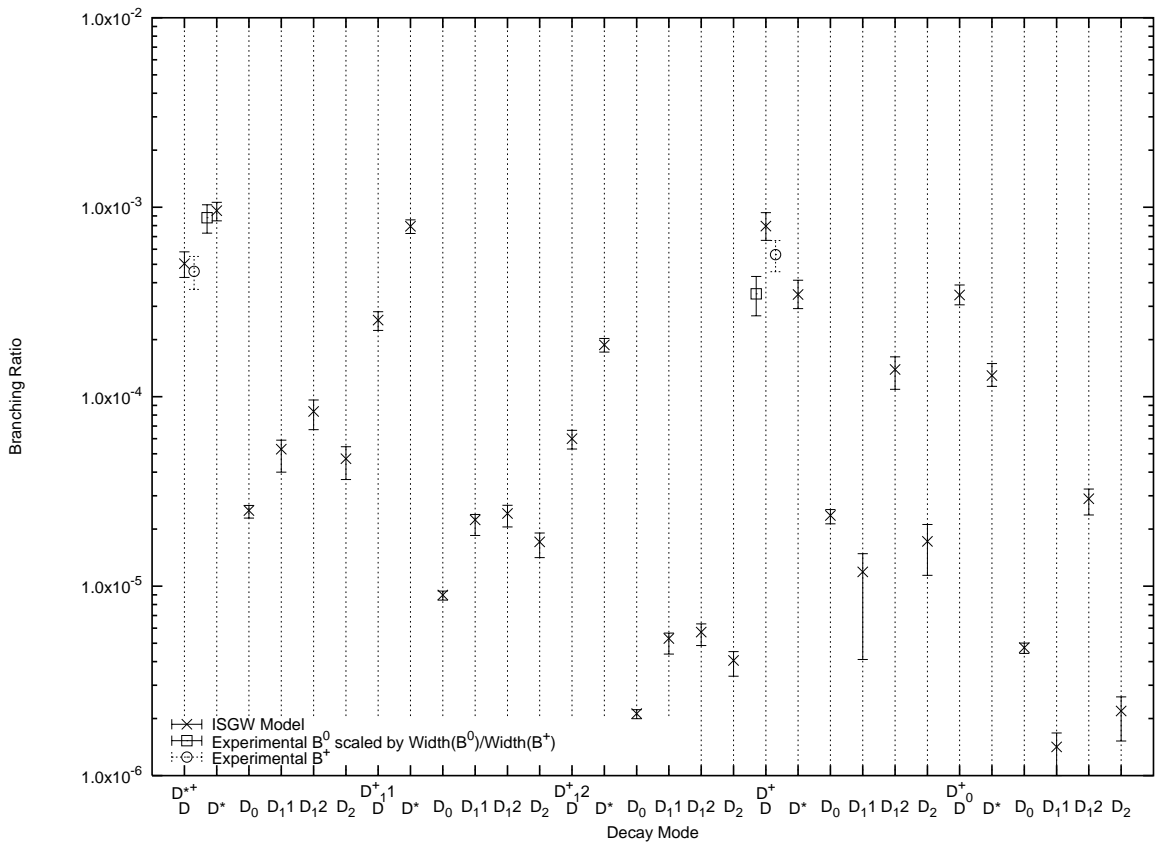}
\caption{$B \rightarrow D D$ Branching Ratios: ISGW model and experimental values.  See the text for references to experimental results.}
\label{fig:DDPlot}
\end{center}
\end{figure}

\begin{figure}[tb]
\begin{center}
\includegraphics[width=11cm,angle=270]{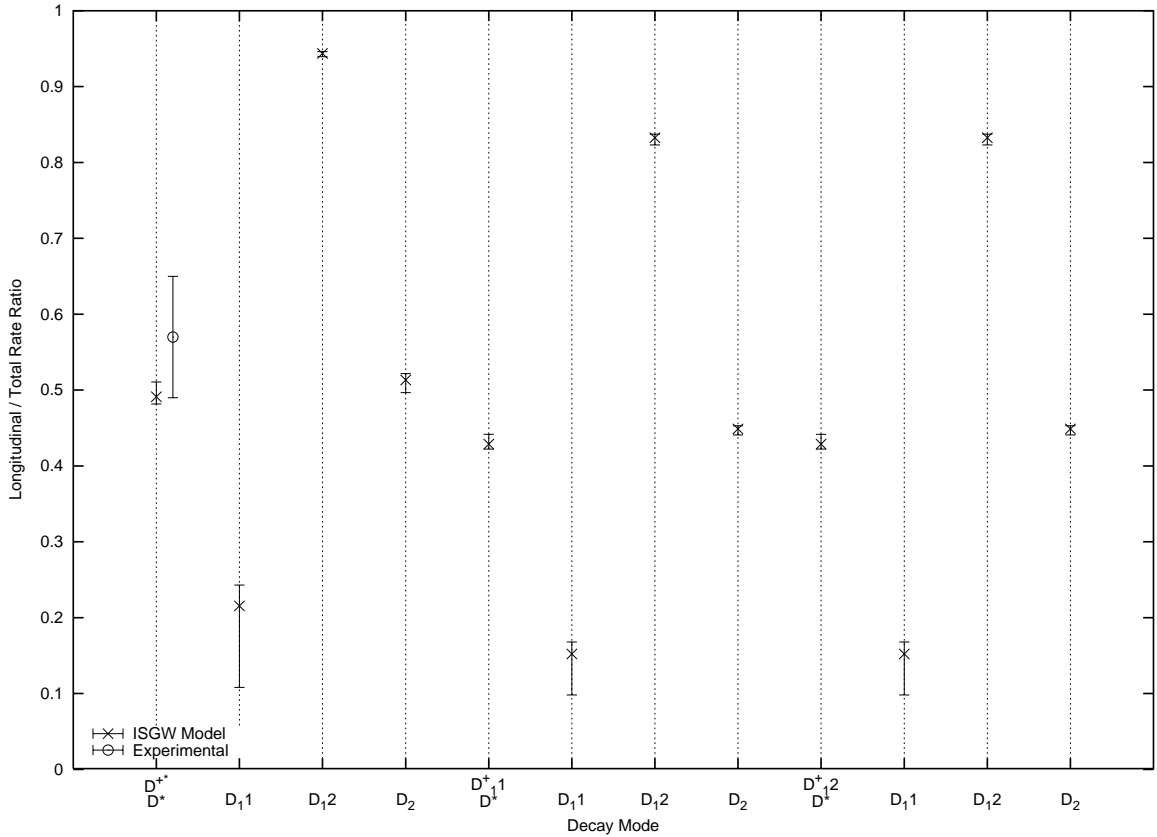}
\caption{Polarization Ratios: ISGW model and experimental values.  See the text for references to experimental results.}
\label{fig:DDPolarisations}
\end{center}
\end{figure}

\end{widetext}

Chen et. al. in ref. \cite{Chen:2005rp} have used generalized factorization to study decays to pseudoscalar 
and/or vector $D$ mesons and compared a number of approaches.  They include penguin and annihilation effects and comment that penguin effects are not negligible in decays to two pseudoscalars.  They have better agreement with experiment in that mode and so this could be a contribution to the discrepancy between our branching ratio to $D$ $D$ and experiment.

\section{Phenomenological Implications for $D_{s0}$ and $D_{s1}$}
\label{sec:PhenomDs0Ds1}

The $D_s(2317)$ ($0^+$) and $D_s(2460)$ ($1^+$) do not fit easily in to the $c\bar{s}$ spectroscopy \cite{Close:2005se}.  These states could be conventional 
$c\bar{s}$ mesons with lower than expected masses \cite{Bardeen:2003kt}.  Alternatively they could be multiquark or molecular mesons associated with $DK$ and 
$D^* K$ thresholds \cite{Barnes:2003dj}.  Measuring branching ratios to $D$ or $D^*$ with $D_s(2317)$ or $D_s(2460)$ would help determine the nature of these 
mesons.  If the measured branching ratios are consistent with our predictions this would support them being conventional mesons.  If this is the case, the 
mixing angle of the $^3P_1$ and $^1P_1$ states in to physical axial vector mesons $D_s(2460)$ and $D_s(2536)$ can also be determined.

Products of branching ratios have been measured for the $D_s(2317)$ and $D_s(2460)$ modes from BABAR \cite{Aubert:2004pw} and Belle \cite{Krokovny:2003zq} and 
included in the PDG Review 2005 Partial Update \cite{PDG05}.  Following Datta and O'Donnell \cite{Datta:2003re} we estimate lower limits for modes containing 
these mesons.  

For decays to $D_s(2317)$ $D$ we get experimental lower limits of $(9.0 \pm 3.2) \times 10^{-4}$ for the $B^{+}$ decay and $(1.1 \pm 0.4) \times 10^{-3}$ for 
the $B^{0}$ decay.  For decays to $D_s(2317)$ $D^{*}$ we obtain $(9 \pm 7) \times 10^{-4}$ and $(1.5 \pm 0.6) \times 10^{-3}$ respectively.  These are 
consistent with the model predictions.

We parameterize the mixing of the axial $D_s$ states in terms of a mixing angle $\phi$.  The physical states are then given by:
\begin{eqnarray}
\left|D_{s1}\right> &=& \sin{\phi} \left|^3P_1\right> + \cos{\phi} \left|^1P_1\right> ,\\
\left|D_{s1}'\right> &=& \cos{\phi} \left|^3P_1\right> - \sin{\phi} \left|^1P_1\right> .
\end{eqnarray}
When $\phi = 0$: $D_{s1}$ is purely $^1P_1$ and corresponds to our $D_{s1}2$, and $D_{s1}'$ is purely $^3P_1$ and corresponds to our $D_{s1}1$.  In the heavy 
quark limit the physical states are $^{3/2}1^+$ and $^{1/2}1^+$ ($^jJ^P$ j-j coupling eigenstates) corresponding to $\phi = -54.7^{\circ}$.

Plots of branching ratios to these mesons and $D$ and $D^{*}$ as a function of $\phi$ are shown in Figures \ref{fig:DsMixingDsD} and \ref{fig:DsMixingDsDv} 
respectively.  We have ignored the mass difference and assumed the $D_{s}(2460)$ to be a conventional scalar meson.  In our conventions the decay constants of 
the $^3P_1$ and $^1P_1$ axials have opposite sign (see Appendix \ref{app:DecayConstants}) and this must be taken in to account when calculating the 
amplitudes.  The estimated experimental limits are shown.  Because only limits exist it is not yet possible to determine the mixing angle using these decays.  
With measured branching ratios these decays provide a relatively clean way to measure the mixing angle but dependent on the $D_{s1}$ decay constants.

\begin{widetext}

\begin{figure}[tb]
\begin{center}
\includegraphics[width=11cm]{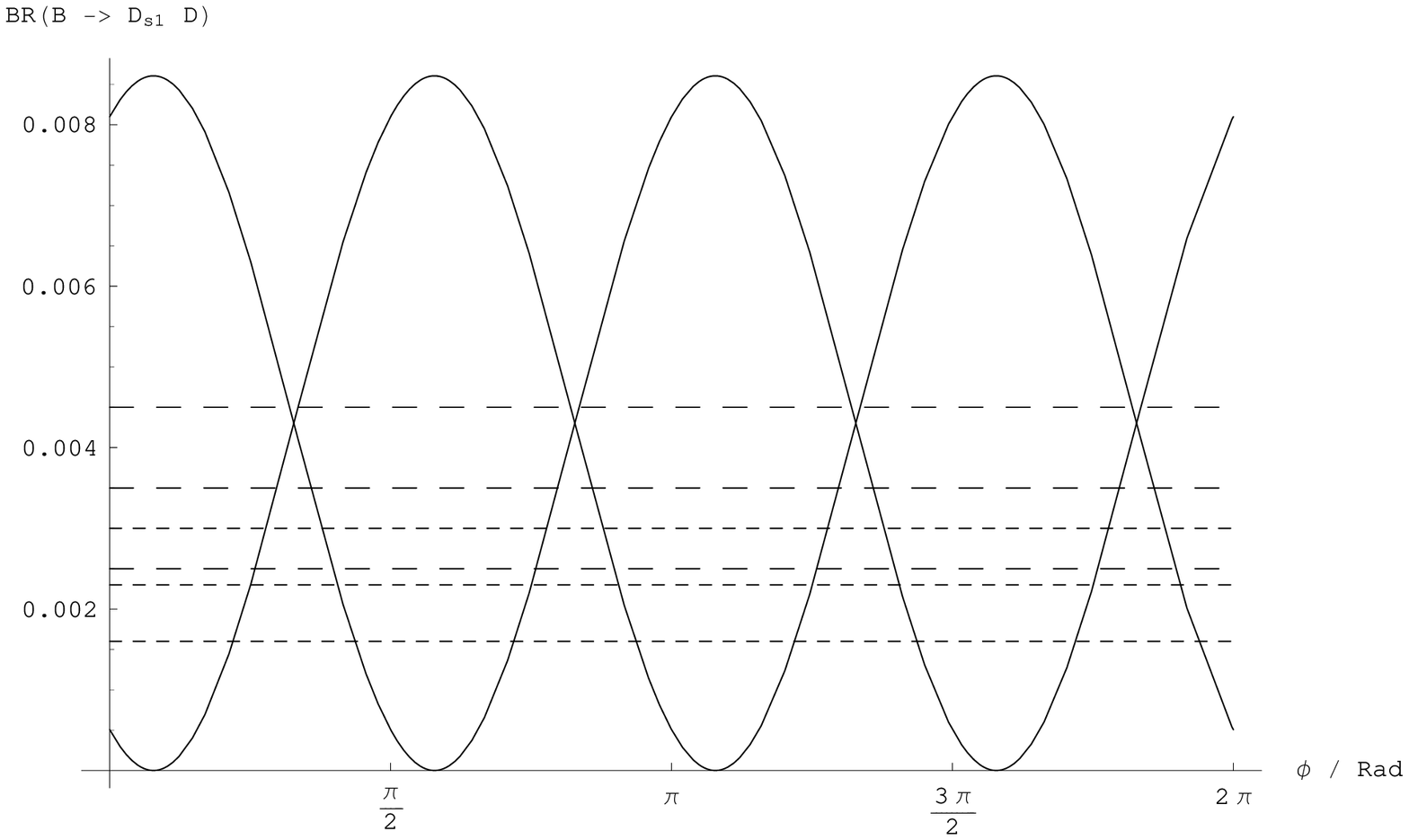}
\caption{Branching Ratios to $D_{s1}$ $D$ as a function of mixing angle.  At $\phi=0$ the top curve is $D_{s1}'$ and the lower curve is $D_{s1}$.  The 
experimental lower limits with $1 \sigma$ errors are shown for $B^+$ (finer dashing) and scaled $B^0$ (coarser dashing).}
\label{fig:DsMixingDsD}
\end{center}
\end{figure}

\begin{figure}[tb]
\begin{center}
\includegraphics[width=11cm]{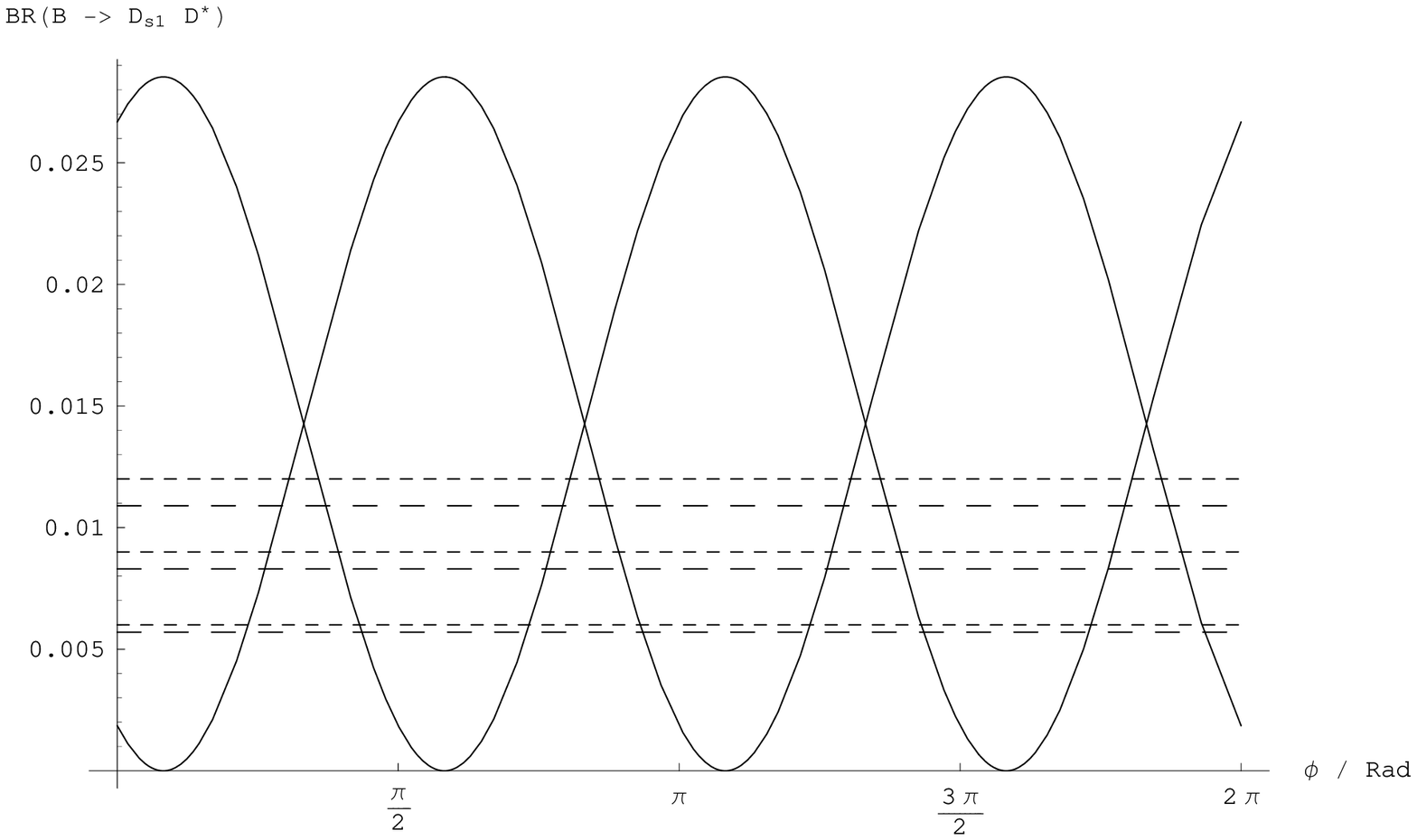}
\caption{Branching Ratios to $D_{s1}$ $D^*$ as a function of mixing angle.  At $\phi=0$ the top curve is $D_{s1}'$ and the lower curve is $D_{s1}$. The 
experimental lower limits with $1 \sigma$ errors are shown for $B^+$ (finer dashing) and scaled $B^0$ (coarser dashing).}
\label{fig:DsMixingDsDv}
\end{center}
\end{figure}

\end{widetext}

Datta and O'Donnell \cite{Datta:2003re} claim that there is a discrepancy between experiment and theory.  However they only consider ratios of decay constants 
and ignore the mass differences between the s and p-wave mesons.  Here we have shown that if the ISGW model is used and the mass differences are taken in to 
account there is no inconsistency between experiment and theory.

Ignoring effects due to the different masses between the $^3P_1$ and $^1P_1$ axial vector mesons, the only difference in branching ratios is due to the decay 
constants.  The quark model expressions for decay constants given in Appendix \ref{app:DecayConstants} imply that the branching ratio to $^3P_1$ $D_s$ is 
larger than that to $^1P_1$ $D_s$ because the first is proportional to the reduced mass whereas the later is proportional to the inverse of the difference of 
inverse masses.  In the equal quark mass limit only the $^3P_1$ meson can be produced, not the $^1P_1$.  In the heavy quark limit only the $^{1/2}1^+$ meson 
can be produced, not the $^{3/2}1^+$ as remarked in Ref. \cite{LeYaouanc:HQSumRules}.

The axials and vectors can not be compared to the scalars and pseudoscalars as simply.  This is because the amplitudes involve different combinations of form factors as seen in Appendix \ref{app:PolAmplitudes}.

\section{Conclusions}
\label{sec:Conclusions}
The ISGW model with the factorization hypothesis produces robust predictions for branching ratios and polarization fractions for the $B \rightarrow D_s D$ decays, which are in general agreement with the limited experimental data.  The predictions are also in line with those from BSW, HQET and pole models.  We have made predictions for all combinations of s and p-wave final state mesons that can be measured in $B$ decays.  There are opportunities to probe the nature of the $D_s(2317)$ and $D_s(2460)$ states and determine the axial vector meson mixing using these decays.  Measuring more polarization fractions will be interesting because they weigh different form factors at the same momentum transfer.  This method has also be used to calculate branching ratios and polarization fractions in $B \rightarrow D D$ decays.

With the B factories at BABAR and Belle, the Tevatron and later LHCb, more precise data and also data on other decay modes should emerge.  

It is important to note that these approximations are robust for the generic class of $D_s$ $D$ decays but are more suspect for color suppressed $B \rightarrow J/\psi K$ (and excited) decays.  These are outside the kinematically valid region and experimental data on these modes disagree with results from this model.  In addition, the factorization hypothesis fails for these decays as evidenced by the observation of a $\chi_{c0}$ $K^+$ decay mode at BABAR \cite{Aubert:2003vc} and Belle \cite{Abe:2001mw}.  A possible contribution to these discrepancies could be final state rescattering of $(c\bar{s})$ $(u\bar{c})$ in to $(c\bar{c})$ $(u\bar{s})$ final states \cite{RescatteringPaper}.

\section*{Acknowledgments}
The initial idea for this work came from Frank Close and I thank Frank Close and Eric Swanson for useful discussions.  I thank Riccardo Faccini, Giuseppe Finocchiaro and Hai-Yang Cheng for comments on an earlier version of this paper.  This work was supported by a studentship from PPARC.

\bibliography{WeakDecays}
\bibliographystyle{h-physrev}

\appendix

\begin{widetext}
\section{General Parameterization of Vector-Axial Currents between Mesons}
\label{app:GeneralV-A}

When X is a pseudoscalar $^1S_0$ $0^{-+}$:
\begin{equation}
\left<X\left|V_{\mu}\right|B\right> \equiv f_+ (P_B + P_X)_{\mu} + f_- (P_B - P_X)_{\mu} .
\end{equation}
\hspace{2ex}

When X is a vector $^3S_1$ $1^{--}$ with polarization vector $\epsilon_{\mu}$:
\begin{equation}
\left<X\left|A_{\mu}\right|B\right> \equiv f \epsilon^*_{\mu} + a_+ (\epsilon^* \cdot P_B)(P_B + P_X)_{\mu} + a_- (\epsilon^* \cdot P_B)(P_B - P_X)_{\mu} ,
\end{equation}
\begin{equation}
\left<X\left|V_{\mu}\right|B\right> \equiv i g \epsilon_{\mu\nu\rho\sigma} \epsilon^{*\nu}(P_B + P_X)^{\rho} (P_B - P_X)^{\sigma} .
\end{equation}
\hspace{2ex}

When X is a scalar $^3P_0$ $0^{++}$:
\begin{equation}
\left<X\left|A_{\mu}\right|B\right> \equiv u_+ (P_B + P_X)_{\mu} + u_- (P_B - P_X)_{\mu} .
\end{equation}
\hspace{2ex}

When X is an axial vector $^3P_1$ $1^{++}$ with polarization vector $\epsilon_{\mu}$:
\begin{equation}
\left<X\left|V_{\mu}\right|B\right> \equiv l \epsilon^*_{\mu} + c_+ (\epsilon^* \cdot P_B)(P_B + P_X)_{\mu} + c_- (\epsilon^* \cdot P_B)(P_B - P_X)_{\mu} ,
\end{equation}
\begin{equation}
\left<X\left|A_{\mu}\right|B\right> \equiv i q \epsilon_{\mu\nu\rho\sigma} \epsilon^{*\nu}(P_B + P_X)^{\rho} (P_B - P_X)^{\sigma} .
\end{equation}
\hspace{2ex}

When X is an axial vector $^1P_1$ $1^{+-}$ with polarization vector $\epsilon_{\mu}$:
\begin{equation}
\left<X\left|V_{\mu}\right|B\right> \equiv r \epsilon^*_{\mu} + s_+ (\epsilon^* \cdot P_B)(P_B + P_X)_{\mu} + s_- (\epsilon^* \cdot P_B)(P_B - P_X)_{\mu} ,
\end{equation}
\begin{equation}
\left<X\left|A_{\mu}\right|B\right> \equiv i v \epsilon_{\mu\nu\rho\sigma} \epsilon^{*\nu}(P_B + P_X)^{\rho} (P_B - P_X)^{\sigma} ,
\end{equation}
\hspace{2ex}

When X is a tensor $^3P_2$ $2^{++}$ with polarization tensor $\epsilon_{\mu\nu}$:
\begin{equation}
\left<X\left|A_{\mu}\right|B\right> \equiv k \epsilon^*_{\mu\nu} P_B^\nu + b_+ (\epsilon^*_{\alpha\beta}P_B^{\alpha}P_B^{\beta})(P_B + P_X)_{\mu} + b_- (\epsilon^*_{\alpha\beta}P_B^{\alpha}P_B^{\beta})(P_B - P_X)_{\mu} ,
\end{equation}
\begin{equation}
\left<X\left|V_{\mu}\right|B\right> \equiv i h \epsilon_{\mu\nu\lambda\rho}\epsilon^{*\nu\alpha}P_{B\alpha}(P_B + P_X)^{\lambda}(P_B - P_X)^{\rho} .
\end{equation}
\hspace{2ex}

\end{widetext}

\section{Polarization Vectors for $J=1$ and $J=2$ mesons}
\label{app:PolVectors}
For a vector meson with 4-momentum $(E,0,0,\mq)$ and mass $M$ the polarization vectors are
\begin{eqnarray} 
\epsilon_+^{\mu} &=& -\frac{1}{\sqrt{2}} (0,1,i,0) ,\\
\epsilon_-^{\mu} &=& \frac{1}{\sqrt{2}} (0,1,-i,0) ,\\
\epsilon_L^{\mu} &=& \frac{1}{M}(\mq,0,0,E) .
\end{eqnarray}

For a vector meson with 4-momentum $(E,0,0,-\mq)$ and mass $M$ the polarization vectors (in the spin rather than the helicity basis) are
\begin{eqnarray} 
\epsilon_+^{\mu} &=& -\frac{1}{\sqrt{2}} (0,1,i,0) ,\\
\epsilon_-^{\mu} &=& \frac{1}{\sqrt{2}} (0,1,-i,0) ,\\
\epsilon_L^{\mu} &=& \frac{1}{M}(-\mq,0,0,E) .
\end{eqnarray}

For a tensor meson with 4-momentum $(E,0,0,\mq)$ and mass $M$ the polarization tensors can be obtained by taking the outer product of two spin-1 polarization vectors with the appropriate Clebsh-Gordon coefficients.  The polarization tensors are given by:
\begin{eqnarray} 
\epsilon_{++}^{\mu\nu} &=&  \frac{1}{2} \left[ \begin{array}{cccc} 0 & 0 & 0 & 0 \\ 0 & 1 & i & 0 \\ 0 & i & -1 & 0 \\ 0 & 0 & 0 & 0 \end{array} \right] ,\\
\epsilon_{+}^{\mu\nu} &=&  -\frac{1}{2M} \left[ \begin{array}{cccc} 0 & \mq & i\mq & 0 \\ \mq & 0 & 0 & E \\ i\mq & 0 & 0 & iE \\ 0 & E & iE & 0 \end{array} \right] ,\\
\epsilon_{0}^{\mu\nu} &=&  \sqrt{\frac{2}{3}} \left[ \begin{array}{cccc} \frac{\mq^2}{M^2} & 0 & 0 & \frac{\mq E}{M^2} \\ 0 & -\frac{1}{2} & 0 & 0 \\ 0 & 0 & -\frac{1}{2} & 0 \\ \frac{\mq E}{M^2} & 0 & 0 & \frac{E^2}{M^2} \end{array} \right] ,\\
\epsilon_{-}^{\mu\nu} &=&  \frac{1}{2M} \left[ \begin{array}{cccc} 0 & \mq & -i\mq & 0 \\ \mq & 0 & 0 & E \\ -i\mq & 0 & 0 & -iE \\ 0 & E & -iE & 0 \end{array} \right] ,\\
\epsilon_{--}^{\mu\nu} &=&  \frac{1}{2} \left[ \begin{array}{cccc} 0 & 0 & 0 & 0 \\ 0 & 1 & -i & 0 \\ 0 & -i & -1 & 0 \\ 0 & 0 & 0 & 0 \end{array} \right] .
\end{eqnarray}

\begin{widetext}

\section{Polarization Amplitudes in Terms of Form Factors}
\label{app:PolAmplitudes}

For a pseudoscalar $B$ decaying to a pseudoscalar (X) and a pseudoscalar (Y) the amplitude is given by
\begin{equation}
A = i f_Y \left[ \left(M_B^2 - M_X^2\right)f_+ + M_Y^2 f_- \right] .
\end{equation}

For a decay in to a vector (X) and a pseudoscalar (Y) only a longitudinal vector is allowed and the amplitude is given by
\begin{equation}
A_L = -i \frac{f_Y \mq M_B}{M_X} \left[f + a_+ (M_B^2 - M_X^2) + a_- M_Y^2 \right] .
\end{equation}

For a decay in to a pseudoscalar (X) and a vector (Y) only a longitudinal vector is allowed and the amplitude is given by
\begin{equation}
A_L = -2 f_Y f_+ \mq M_B .
\end{equation}

For a decay in to a vector (X) and a vector (Y) the polarizations amplitudes for X positive, negative and longitudinal polarization respectively are
\begin{equation}
\label{eqn:polamps:vv:pm}
A_{+-} = - f_Y M_Y \left(f + 2 g \mq M_B\right) ,
\end{equation}
\begin{equation}
\label{eqn:polamps:vv:mp}
A_{-+} = - f_Y M_Y \left(f - 2 g \mq M_B\right) ,
\end{equation}
\begin{eqnarray}
\label{eqn:polamps:vv:ll}
A_{ll} &=& \frac{f_Y}{M_X} \left[ f \left(\mq^2 + \frac{1}{4M_B^2}(M_B^2 + M_X^2 - M_Y^2)(M_B^2 + M_Y^2 - M_X^2)\right) + 2 a_+ \mq^2 M_B^2 \right] ,\\
 &=& \frac{f_Y}{M_X} \left[ f P_Y.P_X + 2 a_+ \mq^2 M_B^2 \right] .
\end{eqnarray}

For a decay in to a tensor (X) and a pseudoscalar (Y) only a longitudinally polarized tensor is allowed and the amplitude is given by
\begin{equation}
A_L = -i f_Y\left[k + b_+ (M_B^2 - M_X^2) + b_- M_Y^2 \right] \sqrt{\frac{2}{3}}\frac{M_B^2 \mq^2}{M_X^2} .
\end{equation}

For a decay in to a tensor (X) and a vector (Y) the tensor can not have polarization $\pm 2$.  The polarization amplitudes for X positive, negative and longitudinally polarized respectively are given by
\begin{equation}
A_{+-} = -f_Y \frac{M_B \mq M_Y}{M_X \sqrt{2}} \left(k + 2 h M_B \mq \right) ,\\
\end{equation}
\begin{equation}
A_{-+} = -f_Y \frac{M_B \mq M_Y}{M_X \sqrt{2}} \left(k - 2 h M_B \mq \right) ,\\
\end{equation}
\begin{equation}
A_{ll} = \sqrt{\frac{2}{3}} \frac{\mq M_B f_Y}{M_X^2} \left[ k \left(\mq^2 + \frac{1}{4M_B^2}(M_B^2 + M_X^2 - M_Y^2)(M_B^2 + M_Y^2 - M_X^2)\right) + 2 M_B^2 \mq^2 b_+ \right] .
\end{equation}

From these expressions it can be seen that the parity violation arises from a non-zero form factors $g$ and $h$.

Amplitudes for when Y is an axial vector or scalar can be obtained simply by using the correct decay constant.  Amplitudes when X is an axial vector or scalar can be obtained by replacing the vector or pseudoscalar form factors by the analogous axial or scalar ones given in Appendix \ref{app:GeneralV-A} \emph{and multiplying by $-1$ to account for the difference in sign of the vector and axial currents in V-A}.

\section{ISGW Form Factors for Harmonic Oscillator Wavefunctions}
\label{app:ISGWFormFactors}

We define:
\begin{equation}
\bbx^2 \equiv \frac{1}{2}(\beta^2_B + \beta^2_X) ,
\end{equation}
\begin{equation}
\mu_{\pm} \equiv \left(\frac{1}{m_q} \pm \frac{1}{m_b}\right)^{-1} ,
\end{equation}
and
\begin{equation}
F_n = \sqrt{\frac{\mmx}{\mmb}} \left(\frac{\beta_B \beta_X}{\bbx^2}\right)^{n/2} \exp\left[-\left(\frac{m_{qi}^2}{4\mmx \mmb}\right)\frac{t_m - t}{\kappa^2\bbx^2}\right] .
\end{equation}

The pseudoscalar ($^1S_0$)  form factors are
\begin{eqnarray}
f_+ =& F_3 \left[1 + \frac{m_b}{2 \mu_-} - \frac{m_b m_q m_{qi} \beta_B^2}{4 \mu_+ \mu_- \mmx \bbx^2} \right] ,\\ 
f_- =& F_3 \left[1 - (\mmx + \mmb)\left(\frac{1}{2 m_q} - \frac{m_{qi} \beta_B^2}{4\mu_+ \mmx \bbx^2} \right) \right] .
\end{eqnarray}

The vector ($^3S_1$) form factors are
\begin{eqnarray}
f =& 2 \mmb F_3 \\
g =& \frac{1}{2} F_3 \left[ \frac{1}{m_q} - \frac{m_{qi} \beta_B^2}{2 \mu_- \mmx \bbx^2}\right] ,\\
a_+ =& - \frac{F_3}{2 \mmx} \left[1 + \frac{m_{qi}}{m_b}\left(\frac{\beta_B^2 - \beta_X^2}{\beta_B^2 + \beta_X^2}\right) - \frac{m_{qi}^2 \beta_X^4}{4 \mu_- \mmb \bbx^4}\right] ,\\
a_- =& \frac{F_3}{2 \mmx} \left[1 - \frac{\mmx}{\mmb} + \frac{\mmx^2}{m_q \mmb} - \frac{\beta_B^2 m_{qi}\mmx}{2\bbx^2 \mmb \mu_+}\right] .
\end{eqnarray}

The axial vector ($^1P_1$) form factors are
\begin{eqnarray}
r =& F_5 \frac{\mmb \beta_B}{\sqrt{2} \mu_+} ,\\
\label{Equ:ISGWFormFactor:v}
v =& F_5 \frac{\beta_B}{4 \sqrt{2} m_b m_q} ,\\
s_+ =& F_5 \frac{m_{qi}}{\sqrt{2} \mmb \beta_B} \left[1 + \frac{m_b}{2 \mu_-} - \frac{m_b m_q m_{qi} \beta_B^2}{4 \mu_+ \mu_- \mmx \bbx^2}\right] ,\\
s_- =& F_5 \frac{m_{qi}}{\sqrt{2} \mmb \beta_B} \left[ 1 - \frac{\mmx + \mmb}{2 m_q} + \frac{(\mmx + \mmb) \beta_B^2 m_{qi}}{4 \mu_+ \bbx^2 \mmx}\right] .
\end{eqnarray}
Note the difference in $v$ compared with B43 of \cite{ISGW}, we have checked this and believe our expression is correct.  We comment on this in Section \ref{sec:ISGW}.

The scalar ($^3P_0$) form factors are
\begin{eqnarray}
\label{Equ:ISGWFormFactor:up}
u_+ =& - F_5 \frac{m_b m_q m_{qi}}{\sqrt{6} \beta_B \mmx \mu_-} ,\\
u_- =& F_5 \frac{m_{qi} (\mmx + \mmb)}{\sqrt{6} \beta_B \mmx} .
\end{eqnarray}
Note the negative sign in $u_+$ compared with that in equation B37 of \cite{ISGW}.  We comment on this in Section \ref{sec:ISGW}.

The axial vector ($^3P_1$) form factors are
\begin{eqnarray}
q =& F_5 \frac{m_{qi}}{2 \mmx \beta_B} , \\
l =& - F_5 \beta_B \mmb \left[ \frac{1}{\mu_-} + \frac{m_{qi} (t_m - t)}{2\mmb \beta_B^2 \kappa^2}\left(\frac{1}{m_q} - \frac{m_{qi}\beta_B^2}{2\bbx\mmx\mu_-}\right)\right] ,\\
c_+ =& F_5 \frac{m_{qi} m_b}{4 \mmb \beta_B \mu_-}\left[ 1 - \frac{m_{qi} m_q \beta_B^2}{2 \mmx \mu_- \bbx^2}\right] ,\\
c_- =& - F_5 \frac{m_{qi}(\mmb + \mmx)}{4 \mmb \beta_B} \left[ \frac{1}{m_q} - \frac{m_{qi} \beta_B^2}{2 \mu_- \bbx^2 \mmx}\right] .
\end{eqnarray}

The tensor ($^3P_2$) form factors are
\begin{eqnarray}
h =& F_5 \frac{m_{qi}}{2\sqrt{2} \mmb \beta_B} \left[ \frac{1}{m_q} - \frac{m_{qi} \beta_B^2}{2 \mmx \mu_- \bbx^2}\right] ,\\
k =& \sqrt{2} F_5 \frac{m_{qi}}{\beta_B} ,\\
b_+ =& - F_5 \frac{m_{qi}}{2 \sqrt{2} \mmx m_b \beta_B} \left[1 - \frac{m_{qi} m_b \beta_X^2}{2 \mu_+ \mmb \bbx^2} + \frac{m_{qi} m_b \beta_X^2}{4 \mmb \mu_- \bbx^2}\left(1 - \frac{m_{qi} \beta_X^2}{2\mmb\bbx^2}\right)\right] ,\\
b_- =& F_5 \frac{m_{qi}}{2\sqrt{2}\mmb^2\beta_B} \left[ -1 + \frac{\mmx}{m_q} + \frac{\mmb}{\mmx} - \frac{m_{qi}}{\mu_+} + \frac{m_{qi}(\mmx + \mmb)}{2 m_b m_q} \right] .
\end{eqnarray}

\section{ISGW Form Factors for General Wavefunctions}
\label{app:ISGWGenFormFactors}

\newcommand{\ov}[1]{\left<#1\right>}

We define:
\begin{equation}
\bbx^2 \equiv \frac{1}{2}(\beta^2_B + \beta^2_X),
\end{equation}
and
\begin{equation}
\mu_{\pm} \equiv \left(\frac{1}{m_q} \pm \frac{1}{m_b}\right)^{-1}
\end{equation}
In the following $\ov{} \equiv \left<X | B \right>$ and $\ov{O} \equiv \left<X\left|O\right|B\right>$.  For example $\ov{p_z} = \int d^3p.p_z.\psi^*(p) \psi(p)$.  For notational convenience we include the common $\sqrt{\mmx/\mmb}$ factors in $\ov{}$ and $\ov{O}$.

The pseudoscalar ($^1S_0$)  form factors are
\begin{eqnarray}
f_+ =& \ov{} \left(1 + \frac{m_b}{2\mu_-}\right) + \frac{\ov{p_z}}{\mq} \frac{m_b m_q}{2 \mu_- \mu_+} ,\\
f_- =& \ov{} \left(1 - \frac{\mmx + \mmb}{2m_q}\right) - \frac{\ov{p_z}}{\mq} \frac{\mmx + \mmb}{2 \mu_+} .
\end{eqnarray}

The vector ($^3S_1$) form factors are
\begin{eqnarray}
g =& \frac{1}{2} \left[ \frac{\ov{}}{m_q} + \frac{\ov{p_z}}{\mq \mu_-} \right] ,\\ 
f =& 2 \ov{} \mmb ,\\
a_+ =& - \frac{1}{2 \mmx} \left[ \ov{}\left(1 + \frac{\mmx}{\mmb} - \frac{\mmx^2}{m_q \mmb}\right) - \frac{\ov{p_z}}{\mq}\frac{\mmx^2}{\mmb\mu_+} \right] ,\\
a_- =&  \frac{1}{2 \mmx} \left[ \ov{}\left(1 - \frac{\mmx}{\mmb} + \frac{\mmx^2}{m_q \mmb}\right) + \frac{\ov{p_z}}{\mq}\frac{\mmx^2}{\mmb\mu_+} \right] .
\end{eqnarray}

The axial vector ($^1P_1$) form factors are
\begin{eqnarray}
r =& \ov{p_+} \frac{\mmb}{\mu_+} ,\\
v =& \ov{p_+} \frac{1}{4 m_b m_q} ,\\
s_+ =& \ov{} \frac{\mmx}{\mq \mmb} \left(1 + \frac{m_b}{2 \mu_-}\right) + \ov{p_z} \frac{m_b m_q \mmx}{2 \mmb \mu_+ \mq^2 \mu_-} - \ov{p_+} \frac{1}{2 \mmb \mu_+} \left(1 + \frac{m_b m_q \mmx}{\mu_- \mq^2} + \frac{m_b m_q}{2 \mu_- \mmx}\right) ,\\
s_- =& \ov{} \frac{\mmx}{\mq \mmb} \left(1 - \frac{(\mmx + \mmb)}{2 m_q}\right) - \ov{p_z} \frac{\mmx (\mmx+\mmb)}{2 \mmb \mu_+ \mq^2} + \ov{p_+} \frac{(\mmx+\mmb)\mmx}{2 \mmb \mu_+ \mq^2} .
\end{eqnarray}

The scalar ($^3P_0$) form factors are
\begin{eqnarray}
u_+ =& - \ov{} \frac{1}{\sqrt{3}}\frac{m_b m_q}{\mq \mu_-} ,\\
u_- =& \ov{} \frac{1}{\sqrt{3}}\frac{(\mmx + \mmb)}{\mq} .
\end{eqnarray}

The axial vector ($^3P_1$) form factors are
\begin{eqnarray}
q =& \ov{} \frac{1}{\sqrt{2} \mq} ,\\
l =& -\frac{\mmb}{\sqrt{2}} \left[ \frac{\ov{p_z}}{\mu_-} + \frac{\ov{} \mq}{m_q} + \frac{\ov{p_+}}{\mu_-} \right] ,\\
c_+ =& \frac{m_b m_q \mmx}{2 \sqrt{2} \mmb \mq^2 \mu_-} \left[ - \frac{\ov{p_+}}{\mu_-} + \frac{\ov{p_z}}{\mu_-} + \frac{\ov{}\mq}{m_q} \right] ,\\
c_- =& \frac{\mmx (\mmx + \mmb)}{2 \sqrt{2} \mmb \mq^2} \left[ \frac{\ov{p_+}}{\mu_-} - \frac{\ov{p_z}}{\mu_-} - \frac{\ov{}\mq}{m_q} \right] .
\end{eqnarray}

The tensor ($^3P_2$) form factors are
\begin{eqnarray}
k =& \ov{} \frac{2 \mmx}{\mq} ,\\
h =& \frac{\mmx}{2 \mmb \mq^2} \left[ \frac{\ov{p_z}}{\mu_-} + \frac{\ov{}\mq}{m_q} - \frac{\ov{p_+}}{\mu_-} \right] ,\\
b_+ =& \frac{\mmx^2}{2 \mq^2 \mmb^2} \left[ \ov{p_z} \left(\frac{1}{\mu_+} + \frac{1}{2 \mu_-}\right) - \ov{p_+} \left(\frac{1}{\mu_+} + \frac{1}{2\mu_-}\right) + \mq \ov{}\left(\frac{1}{m_q} - \frac{2}{\mmx} - \frac{m_b m_q}{\mmx^2 \mu_-}\right) \right] ,\\
b_- =& \frac{\mmx^2}{2 \mq^2 \mmb^2} \left[ \ov{p_z} \left(\frac{1}{\mu_+} - \frac{(\mmx + \mmb)}{2 m_b m_q}\right) - \ov{p_+} \left(\frac{1}{\mu_+} - \frac{(\mmx + \mmb)}{2 m_b m_q}\right) + \mq \ov{}\left(\frac{1}{m_q} - \frac{2}{\mmx} + \frac{(\mmb + \mmx)}{\mmx^2}\right) \right] .
\end{eqnarray}

\end{widetext}

\section{Decay Constants in the Quark Model}
\label{app:DecayConstants}

A non-relativistic quark model calculation using simple harmonic oscillator wavefunctions gives:
\begin{eqnarray} 
f_{^1P_0} &=& i \sqrt{2} \frac{N(Y)}{M_Y} , \\
f_{^1P_1} &=& \sqrt{2} \frac{N(Y)}{M_Y} , \\
f_{^3P_0} &=& -i \sqrt{3} \frac{\beta_Y N(Y)}{\mu_- M_Y}  ,\\
f_{^3P_1} &=& - \sqrt{2} \frac{\beta_Y N(Y)}{\mu_+ M_Y}  ,\\
f_{^1P_1} &=& \frac{\beta_Y N(Y)}{\mu_- M_Y} ,\\
f_{^3P_2} &=& 0 .
\end{eqnarray}
$\beta_Y$ is the SHO wavefunction parameter,
$$N(Y) \propto (4\pi\beta_Y^2)^{\frac{3}{4}} \sqrt{M_Y}$$ and 
$$\frac{1}{\mu_{\pm}} = \frac{1}{m_{q2}} \pm \frac{1}{m_{q1}} .$$
Here $m_{q1}$ is the outgoing quark mass and $m_{q2}$ is the outgoing antiquark mass.  The sign of the $^3P_1$ decay constant depends on which way around the spin and orbital angular momentum are coupled.  We use the convention consistent with the argument in Ref. \cite{LeYaouanc:HQSumRules}.  In this work, a different relative sign here would only lead to the branching ratios shown in Figs. \ref{fig:DsMixingDsD} \& \ref{fig:DsMixingDsDv} being shifted in $\phi$.

\par
For general wavefunctions the above equations are replaced by:
\begin{widetext}
\begin{eqnarray}
f_{^1P_0} &=& i \sqrt{2} \frac{N(Y)}{M_Y} <0,0|0>_L ,\\
f_{^1P_1} &=& \sqrt{2} \frac{N(Y)}{M_Y} <0,0|0>_L ,\\
f_{^3P_0} &=& -i \frac{1}{\sqrt{6}} \frac{N(Y)}{\mu_- M_Y} \left( <1,0|p_z|0>_L + <1,1|p_+|0>_L + <1,-1|p_-|0>_L \right) ,\\
f_{^3P_1} &=& - \frac{1}{2} \frac{N(Y)}{\mu_+ M_Y}  \left( <1,1|p_+|0>_L + <1,-1|p_-|0>_L \right) ,\\
f_{^1P_1} &=& \frac{1}{\sqrt{2}} \frac{N(Y)}{\mu_- M_Y}   <1,0|p_z|0>_L ,\\
f_{^3P_2} &=& 0 .
\end{eqnarray}
\end{widetext}
Here the overlaps $<l,m_l|p_i|0>_L$ only contain the \emph{spatial} integrals; the spin and Clebsh-Gordan factors have already been taken care of.  $p_+ = - \frac{1}{\sqrt{2}}(p_x + i p_y)$ and $p_- = \frac{1}{\sqrt{2}}(p_x - i p_y)$.

\par

Note that a tensor meson can not be produced from an axial-vector current.

In the equal mass limit $\frac{1}{\mu_-} = 0$ and so the scalar or ${^1P_1}$ axial can not be produced.  However, the ${^3P_1}$ axial can be produced. 

In the heavy quark limit where the quark mass $m_{q1} \rightarrow \infty$ and $\mu_- = \mu_+$ both axial vectors can be produced.  It is useful to change from the L-S basis $^{2s+1}L_J$ to the j-j coupling basis $^jJ^P$ where $j$ is the total angular momentum of the light degrees of freedom, $J$ is the total angular momentum and $P$ is the parity.  The transformation is given by \cite{LeYaouanc:HQSumRules}:

\begin{eqnarray}
\left|^1P_1\right> &=& \sqrt{\frac{1}{3}} \left|^{1/2}1^+\right> + \sqrt{\frac{2}{3}} \left|^{3/2}1^+\right> \\
\left|^3P_1\right> &=& -\sqrt{\frac{2}{3}} \left|^{1/2}1^+\right> + \sqrt{\frac{1}{3}} \left|^{3/2}1^+\right>
\end{eqnarray}

\begin{eqnarray}
\left|^{3/2}1^+\right> &=& \sqrt{\frac{2}{3}} \left|^1P_1\right> + \sqrt{\frac{1}{3}} \left|^3P_1\right> \\
\left|^{1/2}1^+\right> &=& \sqrt{\frac{1}{3}} \left|^1P_1\right> - \sqrt{\frac{2}{3}} \left|^3P_1\right> 
\end{eqnarray}

In the heavy quark limit $f_{^3P_1} = - \sqrt{2} f_{^1P_1}$ and so:
\begin{eqnarray}
f_{^{3/2}1^+} &=& 0 ,\\ 
f_{^{1/2}1^+} &=& \sqrt{3} f_{^1P_1} ,\\
f_{^{1/2}0^+} &=& -i \sqrt{3} f_{^1P_1} .
\end{eqnarray}

So the $^{3/2}1^+$ axial can not be produced but the $^{1/2}1^+$ axial can.  Note also that, apart from a phase, the scalar has the same decay constant.  This is a specific realization of the general result given by Le Yaouanc et. al. in Ref. \cite{LeYaouanc:HQSumRules}.

It is important to note that if it was the \emph{antiquark} mass that was large there would be a relative negative sign between $\mu_-$ and $\mu_+$ which would change the negative signs around in the basis change between L-S and j-j coupling.  This means that the basis transformation has different signs in $D_s^+$ and $D_s^-$ mesons.

\section{Tables of Results}
\label{app:NumericalResults}

\begin{table}[htb]
\begin{center}
\begin{tabular}{|c|c|c|c|}
\hline
\textbf{Decay Mode} & \textbf{Branching Ratio} & \textbf{Lower} & \textbf{Upper} \\
\hline
$D_s^*$ $D$ & $  1.49\times 10^{-2}  $ & $ 1.27\times 10^{-2} $ & $  1.70\times 10^{-2} $\\
$D_s^*$ $D^*$  & $   3.18\times 10^{-2} $ & $  2.84\times 10^{-2} $ & $  3.50\times 10^{-2} $\\
$D_s^*$ $D_{0}$ & $   6.97\times 10^{-4} $ & $  6.41\times 10^{-4} $ & $  7.36\times 10^{-4} $\\
$D_s^*$ $D_{1}1$ & $  1.54\times 10^{-3} $ & $  1.20\times 10^{-3} $ & $  1.70\times 10^{-3}$ \\
$D_s^*$ $D_{1}2$ & $ 2.20\times 10^{-3}  $ & $ 1.79\times 10^{-3} $ & $  2.51\times 10^{-3} $\\
$D_s^*$ $D_{2}$ & $ 1.34\times 10^{-3}  $ & $ 1.06\times 10^{-3} $ & $  1.54\times 10^{-3} $\\
\hline
$D_{s1}1$ $D$ &  $ 8.10\times 10^{-3}  $ & $  7.18\times 10^{-3} $ & $   8.92\times 10^{-3} $ \\
$D_{s1}1$ $D^*$ & $  2.67\times 10^{-2}  $ & $  2.45\times 10^{-2} $ & $   2.88\times 10^{-2} $ \\
$D_{s1}1$ $D_{0}$ & $  2.70\times 10^{-4}  $ & $  2.55\times 10^{-4} $ & $   2.84\times 10^{-4} $ \\
$D_{s1}1$ $D_{1}1$ & $  6.82\times 10^{-4} $ & $   5.68\times 10^{-4} $ & $   7.25\times 10^{-4} $ \\
$D_{s1}1$ $D_{1}2$ & $  7.17\times 10^{-4} $ & $   6.13\times 10^{-4} $ & $   7.88\times 10^{-4} $ \\
$D_{s1}1$ $D_{2}$ & $  5.07\times 10^{-4} $ & $   4.21\times 10^{-4} $ & $   5.61\times 10^{-4} $ \\
\hline
$D_{s1}2$ $D$  & $  5.07\times 10^{-4}  $ & $   4.53\times 10^{-4}  $ & $   5.53\times 10^{-4} $ \\
$D_{s1}2$ $D^*$  &   $  1.86\times 10^{-3}  $ & $   1.72\times 10^{-3}  $ & $   2.00\times 10^{-3} $ \\
$D_{s1}2$ $D_{0}$ & $ 1.48\times 10^{-5}  $ & $   1.40\times 10^{-5}  $ & $   1.55\times 10^{-5} $ \\
$D_{s1}2$ $D_{1}1$ & $  3.79\times 10^{-5}  $ & $   3.21\times 10^{-5}  $ & $   4.00\times 10^{-5} $ \\
$D_{s1}2$ $D_{1}2$ & $  3.84\times 10^{-5}  $ & $   3.33\times 10^{-5} $ & $    4.18\times 10^{-5} $ \\
$D_{s1}2$ $D_{2}$ & $  2.61\times 10^{-5}  $ & $   2.19\times 10^{-5}  $ & $   2.87\times 10^{-5} $ \\
\hline
$D_{s}$ $D$ & $  1.76\times 10^{-2}  $ & $ 1.49\times 10^{-2} $ & $  2.05\times 10^{-2} $\\
$D_{s}$ $D^*$ & $ 7.51\times 10^{-3} $ & $  6.38\times 10^{-3} $ & $  8.88\times 10^{-3} $\\
$D_{s}$ $D_{0}$ & $  4.63\times 10^{-4} $ & $  4.21\times 10^{-4} $ & $  4.94\times 10^{-4} $\\
$D_{s}$ $D_{1}1$ & $   2.16\times 10^{-4} $ & $  7.77\times 10^{-5} $ & $  2.67\times 10^{-4}$ \\
$D_{s}$ $D_{1}2$ & $  2.76\times 10^{-3} $ & $  2.19\times 10^{-3} $ & $  3.20\times 10^{-3}$ \\
$D_{s}$ $D_{2}$ & $  3.21\times 10^{-4} $ & $  2.15\times 10^{-4} $ & $  3.92\times 10^{-4} $\\
\hline
$D_{s0}$ $D$  & $ 3.36\times 10^{-3} $ & $  2.97\times 10^{-3} $ & $  3.81\times 10^{-3} $\\
$D_{s0}$ $D^*$ & $  1.28\times 10^{-3} $ & $  1.12\times 10^{-3} $ & $  1.50\times 10^{-3} $\\
$D_{s0}$ $D_{0}$ & $  4.98\times 10^{-5} $ & $  4.64\times 10^{-5} $ & $  5.27\times 10^{-5} $\\
$D_{s0}$ $D_{1}1$ & $   1.57\times 10^{-5} $ & $  7.04\times 10^{-6} $ & $  1.87\times 10^{-5} $\\
$D_{s0}$ $D_{1}2$ & $  3.05\times 10^{-4} $ & $  2.49\times 10^{-4} $ & $  3.44\times 10^{-4}$ \\
$D_{s0}$ $D_{2}$ & $ 2.44\times 10^{-5} $ & $  1.69\times 10^{-5} $ & $  2.91\times 10^{-5} $\\
\hline
\end{tabular}
\end{center}
\caption{$B \rightarrow D_s D$ branching ratios calculated in the ISGW model.  The branching ratio, lower and upper values give the average and range of results dependent on model choices as discussed in Section \protect\ref{sec:ISGW}.}
\label{table:DsDISGWBRs}
\end{table}

\begin{table}[htb]
\begin{center}
\begin{tabular}{|c|c|c|c|}
\hline
\textbf{Decay Mode} & \textbf{Longitudinal Polarization Fraction} & \textbf{Lower} & \textbf{Upper}\\
\hline
$D_s^*$ $D^*$ &	 0.47  &  0.46  &  0.49 \\
$D_s^*$ $D_{1}1$ &  0.19 &   0.10  &  0.22 \\
$D_s^*$ $D_{1}2$ & 0.93  &  0.92 &   0.93 \\
$D_s^*$ $D_{2}$ & 0.50  &  0.48 &   0.50 \\
\hline
$D_{s1}1$ $D^*$ &  0.42 &   0.42  &  0.44 \\
$D_{s1}1$ $D_{1}1$ &  0.15  &  0.10  &  0.16 \\
$D_{s1}1$ $D_{1}2$ &   0.81 &   0.80 &   0.82 \\
$D_{s1}1$ $D_{2}$ &  0.44  &  0.44  &  0.45 \\
\hline
$D_{s1}2$ $D^*$ &   0.41 &   0.41 &   0.43 \\
$D_{s1}2$ $D_{1}1$ &  0.15 &   0.11 &   0.16 \\
$D_{s1}2$ $D_{1}2$ &  0.76  &  0.75  &  0.77 \\
$D_{s1}2$ $D_{2}$ &   0.43 &   0.43 &   0.44 \\
\hline
\end{tabular}
\end{center}
\caption{$B \rightarrow D_s D$ polarization fractions calculated in the ISGW model.  The polarization fraction, lower and upper values give the average and range of results dependent on model choices as discussed in Section \protect\ref{sec:ISGW}.}
\label{table:DsDISGWPolFractions}
\end{table}


\begin{table}[htb]
\begin{center}
\begin{tabular}{|c|c|c|c|}
\hline
\textbf{Decay Mode} & \textbf{Branching Ratio} & \textbf{Lower} & \textbf{Upper} \\
\hline
$D^{*+}$ $D$ & $   5.04\times 10^{-4} $ & $  4.26\times 10^{-4}  $ & $  5.81\times 10^{-4} $ \\
$D^{*+}$ $D^*$  & $ 9.58\times 10^{-4}  $ & $ 8.48\times 10^{-4}  $ & $  1.06\times 10^{-3} $ \\
$D^{*+}$ $D_{0}$ & $ 2.51\times 10^{-5} $ & $  2.29\times 10^{-5}  $ & $  2.67\times 10^{-5} $ \\
$D^{*+}$ $D_{1}1$ & $  5.29\times 10^{-5} $ & $  4.00\times 10^{-5}  $ & $  5.90\times 10^{-5} $ \\
$D^{*+}$ $D_{1}2$ & $ 8.35\times 10^{-5} $ & $  6.71\times 10^{-5}  $ & $  9.61\times 10^{-5} $ \\
$D^{*+}$ $D_{2}$ & $  4.70\times 10^{-5} $ & $  3.66\times 10^{-5}  $ & $  5.45\times 10^{-5} $ \\
\hline
$D^+_{1}1$ $D$ &  $  2.54\times 10^{-4}  $ & $ 2.24\times 10^{-4}  $ & $  2.81\times 10^{-4} $ \\
$D^+_{1}1$ $D^*$ & $   7.93\times 10^{-4}  $ & $ 7.27\times 10^{-4}  $ & $  8.58\times 10^{-4} $ \\
$D^+_{1}1$ $D_{0}$ & $  8.98\times 10^{-6} $ & $  8.45\times 10^{-6}  $ & $  9.43\times 10^{-6} $ \\
$D^+_{1}1$ $D_{1}1$ & $   2.24\times 10^{-5} $ & $  1.85\times 10^{-5}  $ & $  2.39\times 10^{-5} $ \\
$D^+_{1}1$ $D_{1}2$ & $ 2.42\times 10^{-5} $ & $  2.06\times 10^{-5}  $ & $  2.67\times 10^{-5} $ \\
$D^+_{1}1$ $D_{2}$ & $ 1.71\times 10^{-5}$ & $   1.42\times 10^{-5}  $ & $  1.91\times 10^{-5} $ \\
\hline
$D^+_{1}2$ $D$  & $  6.00\times 10^{-5} $ & $  5.30\times 10^{-5} $ & $  6.64\times 10^{-5} $ \\
$D^+_{1}2$ $D^*$  & $ 1.88\times 10^{-4} $ & $  1.72\times 10^{-4} $ & $  2.03\times 10^{-4} $ \\
$D^+_{1}2$ $D_{0}$ & $ 2.12\times 10^{-6} $ & $  2.00\times 10^{-6} $ & $  2.23\times 10^{-6} $ \\
$D^+_{1}2$ $D_{1}1$ & $  5.30\times 10^{-6} $ & $   4.38\times 10^{-6} $ & $  5.66\times 10^{-6} $ \\
$D^+_{1}2$ $D_{1}2$ & $   5.72\times 10^{-6} $ & $  4.86\times 10^{-6} $ & $  6.32\times 10^{-6} $ \\
$D^+_{1}2$ $D_{2}$ & $  4.05\times 10^{-6} $ & $  3.35\times 10^{-6} $ & $  4.51\times 10^{-6} $ \\
\hline
$D^+$ $D$ & $ 7.95\times 10^{-4}  $ & $ 6.69\times 10^{-4} $ & $  9.36\times 10^{-4} $ \\
$D^+$ $D^*$ & $  3.46\times 10^{-4} $ & $  2.92\times 10^{-4} $ & $  4.12\times 10^{-4} $ \\
$D^+$ $D_{0}$ & $ 2.36\times 10^{-5}  $ & $ 2.13\times 10^{-5} $ & $  2.54\times 10^{-5} $ \\
$D^+$ $D_{1}1$ & $ 1.19\times 10^{-5}  $ & $ 4.10\times 10^{-6} $ & $  1.48\times 10^{-5} $ \\
$D^+$ $D_{1}2$ & $ 1.39\times 10^{-4}  $ & $ 1.09\times 10^{-4} $ & $  1.62\times 10^{-4} $ \\
$D^+$ $D_{2}$ & $   1.72\times 10^{-5} $ & $  1.14\times 10^{-5} $ & $  2.12\times 10^{-5} $ \\
\hline
$D^+_{0}$ $D$ & $  3.44\times 10^{-4}  $ & $ 3.05\times 10^{-4} $ & $  3.88\times 10^{-4} $ \\
$D^+_{0}$ $D^*$ & $ 1.29\times 10^{-4} $ & $  1.13\times 10^{-4} $ & $  1.50\times 10^{-4} $ \\
$D^+_{0}$ $D_{0}$ & $  4.73\times 10^{-6} $ & $  4.42\times 10^{-6} $ & $  5.01\times 10^{-6} $ \\
$D^+_{0}$ $D_{1}1$ & $ 1.42\times 10^{-6} $ & $  6.55\times 10^{-7} $ & $  1.68\times 10^{-6} $ \\
$D^+_{0}$ $D_{1}2$ & $  2.89\times 10^{-5} $ & $  2.37\times 10^{-5} $ & $  3.25\times 10^{-5} $ \\
$D^+_{0}$ $D_{2}$ & $ 2.19\times 10^{-6} $ & $  1.52\times 10^{-6} $ & $  2.60\times 10^{-6} $ \\
\hline
\end{tabular}
\end{center}
\caption{$B \rightarrow D D$ branching ratios calculated in the ISGW model.  The branching ratio, lower and upper values give the average and range of results dependent on model choices as discussed in Section \protect\ref{sec:ISGW}.}
\label{table:DDISGWBRs}
\end{table}

\begin{table}[htb]
\begin{center}
\begin{tabular}{|c|c|c|c|}
\hline
\textbf{Decay Mode} & \textbf{Longitudinal Polarization Fraction} & \textbf{Lower} & \textbf{Upper}\\
\hline
$D^{*+}$ $D^*$ &  $  0.49  $ & $  0.48  $ & $  0.51 $ \\
$D^{*+}$ $D_{1}1$ & $ 0.22  $ & $  0.11  $ & $  0.24 $ \\
$D^{*+}$ $D_{1}2$ & $  0.94  $ & $  0.94  $ & $  0.95 $ \\
$D^{*+}$ $D_{2}$ & $  0.51  $ & $  0.50  $ & $  0.52 $ \\
\hline
$D^+_{1}1$ $D^*$ & $ 0.43 $ & $   0.42  $ & $  0.44 $ \\
$D^+_{1}1$ $D_{1}1$ & $ 0.15 $ & $   0.098  $ & $  0.17 $ \\
$D^+_{1}1$ $D_{1}2$ & $ 0.83 $ & $   0.82  $ & $  0.84 $ \\
$D^+_{1}1$ $D_{2}$ &  $ 0.45  $ & $  0.44  $ & $  0.45 $ \\
\hline
$D^+_{1}2$ $D^*$ &  $  0.43  $ & $  0.42  $ & $  0.44 $ \\
$D^+_{1}2$ $D_{1}1$ & $ 0.15  $ & $  0.098  $ & $  0.17 $ \\
$D^+_{1}2$ $D_{1}2$ & $ 0.83  $ & $  0.82  $ & $  0.84 $ \\
$D^+_{1}2$ $D_{2}$ & $ 0.45  $ & $  0.44  $ & $  0.45 $ \\
\hline
\end{tabular}
\end{center}
\caption{$B \rightarrow D D$ polarization fractions calculated in the ISGW model.  The polarization fraction, lower and upper values give the average and range of results dependent on model choices as discussed in Section \protect\ref{sec:ISGW}.}
\label{table:DDISGWPolFractions}
\end{table}

\begin{table}[htb]
\begin{center}
\begin{tabular}{|c|c|c|c|}
\hline
\textbf{Decay Mode} & \textbf{CP Odd Fraction, $R_{\perp}$} & \textbf{Lower} & \textbf{Upper}\\
\hline
$D^{*+}$ $D^*$ & $  8.61\times 10^{-2} $ & $  7.93\times 10^{-2} $ & $   8.95\times 10^{-2} $ \\
$D^{*+}$ $D_{1}1$ & $ 4.55\times 10^{-1} $ & $   3.99\times 10^{-1} $ & $   6.15\times 10^{-1}$ \\
$D^{*+}$ $D_{1}2$ & $ 5.76\times 10^{-4} $ & $   5.69\times 10^{-4} $ & $   5.89\times 10^{-4}$ \\
$D^{*+}$ $D_{2}$ & $  5.82\times 10^{-2} $ & $   5.68\times 10^{-2} $ & $   5.90\times 10^{-2} $ \\
\hline
$D^+_{1}1$ $D^*$ & $  7.12\times 10^{-2} $ & $   6.65\times 10^{-2} $ & $   7.36\times 10^{-2}$ \\
$D^+_{1}1$ $D_{1}1$ & $  5.00\times 10^{-1} $ & $   4.56\times 10^{-1} $ & $   6.05\times 10^{-1}$ \\
$D^+_{1}1$ $D_{1}2$ & $  9.33\times 10^{-4} $ & $   9.32\times 10^{-4} $ & $   9.35\times 10^{-4}$ \\
$D^+_{1}1$ $D_{2}$ & $ 3.62\times 10^{-2} $ & $   3.45\times 10^{-2} $ & $   3.71\times 10^{-2} $ \\
\hline
$D^+_{1}2$ $D^*$ & $  7.12\times 10^{-2} $ & $   6.65\times 10^{-2} $ & $   7.36\times 10^{-2} $  \\
$D^+_{1}2$ $D_{1}1$ & $ 5.00\times 10^{-1} $ & $   4.56\times 10^{-1} $ & $   6.05\times 10^{-1} $ \\
$D^+_{1}2$ $D_{1}2$ & $ 9.33\times 10^{-4} $ & $   9.32\times 10^{-4} $ & $   9.35\times 10^{-4}$ \\
$D^+_{1}2$ $D_{2}$ & $ 3.62\times 10^{-2} $ & $   3.45\times 10^{-2} $ & $   3.71\times 10^{-2}$ \\
\hline
\end{tabular}
\end{center}
\caption{$B \rightarrow D D$ CP odd fractions $R_{\perp}$ calculated in the ISGW model.  The polarization fraction, lower and upper values give the average and range of results dependent on model choices as discussed in Section \protect\ref{sec:ISGW}.}
\label{table:DDISGWTransFractions}
\end{table}

\end{document}